\documentclass[10pt,onecolumn,aps,prd,preprintnumbers,showpacs,superscriptaddress,nofootinbib,amsmath,amssymb,floats,floatfix,showkeys,notitlepage,longbibliography]{revtex4-2}

\usepackage{orcidlink}
\usepackage{comment}
\usepackage{lipsum}
\usepackage{graphicx}
\usepackage{subfigure}
\usepackage{palatino}
\usepackage{sans}
\usepackage{hyperref}
\hypersetup{colorlinks=true,linkcolor=blue,urlcolor=blue,citecolor=blue}
\usepackage[toc,page]{appendix}
\usepackage[normalem]{ulem}
\usepackage{adjustbox}
\usepackage{latexsym}
\usepackage{amsmath}
\numberwithin{equation}{section}
\usepackage{amssymb}
\usepackage{amsfonts}
\usepackage{dcolumn}
\usepackage{bm}
\usepackage{tikz}
\usetikzlibrary{decorations.pathmorphing}
\usepackage{bigints}
\usepackage{array,tabularx,multirow,booktabs}
\usepackage[tracking=true]{microtype}
\usepackage{soul} %for highlighting
\SetTracking{}{500}
\SetTracking{encoding={*}, shape=sc}{40}
\UseRawInputEncoding %for inputenc error%
\allowdisplaybreaks
\usepackage{mathrsfs}

\usepackage[utf8]{inputenc}
\usepackage{xcolor} % For colored text if needed

\begin{document} \sloppy

\title{Ringdown modulation of acceleration radiation in the Schwarzschild background}

\author{Reggie C. Pantig \orcidlink{0000-0002-3101-8591}} 
\email{rcpantig@mapua.edu.ph}
\affiliation{Physics Department, School of Foundational Studies and Education, Map\'ua University, 658 Muralla St., Intramuros, Manila 1002  Philippines.}

\begin{abstract}
We derive an analytic first-order description of how Schwarzschild ringdown affects a detector-based detailed-balance diagnostic in a near-horizon, single-mode setting. A freely falling two-level system couples to a cavity-filtered outgoing mode of fixed asymptotic frequency, whose static Schwarzschild response gives geometric photon statistics and a detailed-balance ratio governed by the surface gravity. We perturb this baseline by an even-parity, axisymmetric quadrupolar quasinormal mode and work in ingoing Eddington-Finkelstein coordinates, regular at the future horizon. The perturbation shifts the outgoing eikonal through the double-null contraction of the metric perturbation along the outgoing congruence. After fixing the residual endpoint phase calibration on the cavity worldtube, this redshift-map deformation induces a first-order decaying-oscillatory correction to the detector detailed-balance exponent at the quasinormal frequency. We express the geometric response through a closed boundary formula at the sampling radius and state the adiabatic, narrowband, and linear-response conditions under which the result applies. Detector details, including the gap, switching, and wavepacket profile, enter only through a smooth prefactor, while the ringdown dependence is carried by the quasinormal frequency and calibrated response coefficient. The modulation vanishes in the zero-amplitude, late-time, and stationary quadrupolar limits. The result is not a modification of the Hawking temperature, global Hawking flux, or dynamical horizon thermality, but a controlled correction to an operational detector/cavity detailed-balance observable.
\end{abstract}


\maketitle

\section{Introduction}\label{sec1}
Black-hole horizons are associated with thermal behavior in several related
but distinct senses: Hawking radiation at future null infinity
\cite{Hawking:1974rv,Hawking:1975vcx,Hawking_1979,Gibbons:1994cg},
the near-horizon Rindler structure
\cite{Davies:1978zz,Unruh:1976db,Wald:1999vt,Crispino:2007eb}, and
KMS/detailed-balance relations for suitable quantum probes
\cite{Kubo:1957mj,Martin:1959jp,Takagi:1986kn,Sewell:1982zz,Kay:1988mu}.
These notions are often discussed under the common language of horizon
thermality, but they are not identical observables. In particular, a
detector-based detailed-balance relation probes an operational response
function associated with a specified detector trajectory, switching protocol,
field mode, and measurement procedure. This distinction is important in the
present work.

We study how black-hole ringdown affects such an operational
detailed-balance diagnostic. The static reference point is the
Schwarzschild setup considered by Scully and collaborators
\cite{Scully:2017utk}, in which a freely falling two-level system couples to
a selected outgoing cavity mode of fixed asymptotic frequency $\nu$. In that
stationary problem, the excitation and absorption probabilities yield a
geometric photon distribution, and the ratio
$\Gamma_{\rm abs}/\Gamma_{\rm exc}$ is governed by the Schwarzschild surface
gravity. This result is not a statement about the full Hawking flux by
itself; it is a single-mode detector/cavity detailed-balance statement.

Real black holes, however, are not exactly stationary after formation or
perturbation. At late times they relax through damped quasinormal modes
(QNMs) \cite{Vishveshwara:1970zz,Zerilli:1970wzz,Teukolsky:1973ha,
Leaver:1986vnb,Kokkotas:1999bd,Berti:2009kk,Konoplya:2011qq,
LIGOScientific:2016aoc,Konoplya:2022pbc,Konoplya:2023hqb,Cheung:2022rbm}.
It is therefore natural to ask a more limited and operational question:
within the same type of single-mode detector/cavity framework, does the
static detailed-balance exponent remain fixed during ringdown, or does the
time-dependent geometry induce a controlled correction?

In this paper we answer this question analytically at first order in the
ringdown amplitude. We perturb the Schwarzschild background by an
even-parity, axisymmetric quadrupolar QNM and work in ingoing
Eddington-Finkelstein coordinates, so that regularity at the future horizon
is manifest \cite{Eddington:1924pmh,Finkelstein:1958zz}. The outgoing mode
phase is described by an eikonal variable
\[
    u=u_0+\varepsilon\,\delta u ,
\]
where $u_0$ is the Schwarzschild retarded time and $\delta u$ is sourced by
the double-null contraction $h_{ab}k^ak^b$ of the metric perturbation along
the outgoing congruence. After fixing the residual endpoint phase
calibration on the cavity worldtube, this redshift-map deformation produces
a first-order, decaying-oscillatory correction to the detector-inferred
detailed-balance exponent. The carrier frequency and damping rate are the
QNM pair $(\omega_R,\omega_I)$, while the radial dependence is encoded in a
single EF-regular response coefficient $\mathcal{C}_{20}(r_c)$.

The main result may be summarized schematically as
\[
    \ln\frac{\Gamma_{\rm abs}}{\Gamma_{\rm exc}}
    =
    \frac{2\pi\nu}{\kappa}
    +
    \delta_{\rm RD}(v_c),
\]
where the first term is the static Schwarzschild detailed-balance exponent
and the correction has the QNM form
\[
    \delta_{\rm RD}(v_c)
    \propto
    \varepsilon\,
    |\mathcal{C}_{20}(r_c)|\,
    e^{-\omega_I v_c}
    \mathcal{S}\!\left(\omega_R v_c-\arg\mathcal{C}_{20}(r_c)\right).
\]
The precise expression, including the detector-dependent prefactor and the
adiabatic error terms, is given in Theorem 1. The correction vanishes in the
zero-amplitude, late-time, and stationary quadrupolar limits, so the static
Schwarzschild relation is recovered.

We emphasize the scope of this result. We do not claim that the Hawking
temperature, the global Hawking flux, or a general notion of dynamical
horizon thermality is itself modulated by ringdown. What is derived here is
a correction to a particular operational observable: the detailed-balance
ratio of a freely falling two-level detector coupled to a cavity-filtered
outgoing mode in a perturbatively time-dependent Schwarzschild background.
The result is therefore conditional on the single-mode projection, the
adiabatic transit window, the near-horizon eikonal approximation, and the
retarded-phase calibration on the sampling worldtube.

This qualification also clarifies what should, and should not, be regarded as
generic. The appearance of the QNM template---oscillation at \(\omega_R\),
decay at \(\omega_I\), and disappearance in the late-time limit---is a
generic consequence of linear ringdown entering a phase-sensitive
near-horizon observable through the redshift map. By contrast, the numerical
size, sign, and absolute phase of the modulation are not universal: they
depend on the selected outgoing mode, the detector gap and switching
function, the cavity bandwidth, the sampling radius, and the endpoint phase
calibration used to compare the cavity mode with the asymptotic retarded-time
phase. Thus the present calculation should be viewed as a controlled
detector-response probe of time-dependent horizon physics, rather than as a
definition of a dynamical Hawking temperature.

Several related lines of work motivate this question. Unruh-DeWitt detectors
in black-hole spacetimes have long provided operational probes of
near-horizon quantum behavior \cite{DeWitt_1979,Louko:2006zv,
Hodgkinson:2012mr,Tjoa:2022oxv,Conroy:2021aow}. In the quantum-optics
language, atoms crossing cavities near black holes offer a single-mode
realization of horizon-brightened acceleration radiation
\cite{Scully:2017utk,Svidzinsky:2018jkp,Ben-Benjamin:2019opz,
Azizi:2020gff,Azizi:2021qcu,Azizi:2021yto,Camblong:2020pme,
Sen:2022cdx,Sen:2022tru,Das:2023rwg,Das:2025rzz,Jana:2024fhx,
Jana:2025hfl,Bukhari:2022wyx,Bukhari:2023yuy,Ovgun:2025isv,
Pantig:2025okn,Pantig:2025igg}. Other detector-centered studies have
clarified finite-time thermality tests and KMS diagnostics in curved
spacetimes and slowly varying settings
\cite{Dalui:2020qpt,Kaplanek:2021sbo,Shallue:2025zto,Clarke:2024lwi,
Carullo:2025oms}. By contrast, classical ringdown physics is usually
formulated in terms of waveform dynamics rather than detector
detailed-balance observables. The present work connects these two
viewpoints in a controlled first-order calculation.

The paper is organized as follows. Section \ref{sec2} reviews the static
Schwarzschild detector/cavity baseline and introduces the even-parity
ringdown geometry. Section \ref{sec3} derives the EF-regular redshift-map
correction and its effect on the single-mode detector response.
Section \ref{sec4} presents the ringdown-modulated detailed-balance law, the
closed boundary expression for $\mathcal{C}_{20}(r_c)$, and the static-limit
check. Section \ref{sec5} states the regime of validity, including the
near-horizon, adiabatic, gauge-calibration, and Hadamard requirements.
Section \ref{sec6} summarizes the result and discusses possible extensions,
with analog or experimental implications treated only as prospective
applications rather than as established consequences. Unless otherwise
stated, $G=c=\hbar=1$. We use \(\nu\) for the asymptotic frequency of the selected outgoing cavity
mode and \(\omega_A\) for the detector gap. The QNM frequency is denoted
\(\omega=\omega_R-i\omega_I\), with \(\omega_I>0\). The sampling radius of
the cavity is \(r_c\), the corresponding advanced time is \(v_c\), and
\(\mathcal{C}_{20}(r_c)\) denotes the calibrated radial response coefficient
of the quadrupolar perturbation. The small parameter \(\varepsilon\) tracks
the ringdown amplitude.

\section{Background and Setup} \label{sec2}
In this section, we assemble the ingredients used throughout the paper. We first summarize the static Schwarzschild baseline, which is the setting in which a freely falling two-level system couples to a selected outgoing field mode and exhibits thermal detailed balance governed by the surface gravity. This provides the operational statement of near-horizon thermality that our ringdown analysis will perturb. We fix conventions, coordinates, and detector-field coupling so that the subsequent quasinormal-mode (QNM) calculation can be presented as a controlled first-order deformation of these formulas.

\subsection{Static Schwarzschild baseline} \label{ssec2.1}
We work with the Schwarzschild metric \cite{Schwarzschild:1916uq,Wald:1984rg,Carroll_2019,Hollands:2014eia}
\begin{equation}
    ds^{2}= -\left(1-\frac{2M}{r}\right)dt^{2}+\left(1-\frac{2M}{r}\right)^{-1}dr^{2}+r^{2}d\Omega^{2}, \label{2.1}
\end{equation}
where $d\Omega^2 = d\theta^2 + \sin^2\theta\, d\phi^2$ and introduce the tortoise coordinate and Eddington-Finkelstein null coordinates
\begin{equation}
    r_{*}=r+2M\ln \Big|\frac{r}{2M}-1\Big| \quad u=t-r_{*},\quad v=t+r_{*}. \label{2.2}
\end{equation}
The surface gravity is \cite{Bardeen:1973tla,Wald:1984rg}
\begin{equation}
    \kappa=\frac{1}{4M}=\frac{1}{2r_g}, \label{2.3}
\end{equation}
with gravitational radius \(r_g=2M\). Near the horizon \((r\to2M)\), the metric is Rindler-like and \(u\) plays the role of Rindler time for outgoing modes.

Following the operational setup of \cite{Scully:2017utk}, we consider identical two-level systems (gap \(\omega>0\)) that fall freely from rest at infinity along radial geodesics. For such geodesics (specific energy \(E=1\)), the worldline obeys
\begin{equation}
    \frac{dr}{d\tau}=-\sqrt{\frac{2M}{r}},\quad \frac{dt}{d\tau}=\left(1-\frac{2M}{r}\right)^{-1}, \label{2.4}
\end{equation}
so that near the horizon, the outgoing null coordinate pulled back to the worldline has the universal logarithmic form
\begin{equation}
    u(\tau)=u_{0}-\frac{1}{\kappa}\ln \left[\kappa(\tau_{H}-\tau)\right]+\mathcal{O}(\tau_{H}-\tau), \label{2.5}
\end{equation}
where \(\tau\) is the atom's proper time and \(\tau_H\) is the finite proper time at which the horizon is crossed.

We select a single outgoing field mode of frequency \(\nu>0\) defined with respect to \(u\) at future null infinity. The detailed-balance exponent in \(\Gamma_{\mathrm{abs}}/\Gamma_{\mathrm{exc}}\) is governed by this mode frequency \(\nu\) (set operationally by the cavity), while the detector gap, which is denoted \(\omega\) in this subsection and \(\omega_{A}\) elsewhere, which affects only smooth prefactors that cancel in the ratio. In the rotating-wave/anti-rotating-wave decomposition along the worldline, the relevant interaction picture matrix elements carry the phase
\begin{equation}
    \Phi(\tau)=\nu u(\tau)\pm \omega \tau, \label{2.6}
\end{equation}
with \(+\) for excitation with emission (counter-rotating process) and \(-\) for de-excitation with absorption.

To leading order in the atom-field coupling \(g\), the excitation probability for emitting one quantum into the selected mode is
\begin{equation}
    P_{\mathrm{exc}}=g^{2}\Big|\int d\tau e^{i\nu u(\tau)}e^{i\omega \tau}\Big|^{2}. \label{2.7}
\end{equation}
Using the near-horizon form \eqref{2.5}, the integral reduces (via a change of variables \(x\propto \tau_{H}-\tau)\) to a standard gamma-function integral, yielding a Planckian factor \cite{Scully:2017utk},
\begin{equation}
    P_{\mathrm{exc}}=\mathcal{N} \frac{1}{e^{2\pi\nu/\kappa}-1},\quad P_{\mathrm{abs}}=\mathcal{N} \frac{e^{2\pi\nu/\kappa}}{e^{2\pi\nu/\kappa}-1}, \label{2.8}
\end{equation}
where \(\mathcal{N}\) is the same smooth prefactor in both channels (its explicit form depends on \(\omega\) and the long-time windowing but cancels in ratios in the regime \(\omega\gg \nu\) emphasized in Ref. \cite{Scully:2017utk}). Consequently,
\begin{equation}
    \frac{\Gamma_{\mathrm{abs}}}{\Gamma_{\mathrm{exc}}}=e^{2\pi\nu/\kappa} = e^{4\pi r_g \nu}, \label{2.9}
\end{equation}
which is the detailed-balance relation characteristic of a KMS state at local Tolman temperature \(T_{H}=\kappa/2\pi\) when viewed by the freely falling detector sampling the outgoing mode.

Embedding the atoms in a weakly leaky single-mode cavity (as in Ref. \cite{Scully:2017utk}) and iterating \eqref{2.8} leads to a geometric steady state for the mode occupation,
\begin{equation}
    p_{n}=(1-e^{-2\xi}) e^{-2\xi n},\quad
    2\xi=\ln \left(\frac{\Gamma_{\mathrm{abs}}}{\Gamma_{\mathrm{exc}}}\right)=\frac{2\pi\nu}{\kappa}. \label{2.10}
\end{equation}
Eqs. \eqref{2.8}-\eqref{2.10} constitute the baseline we shall perturb in Section \ref{sec3}: during ringdown, \(u(\tau)\) acquires a small, explicitly computable correction, and the exponent \(2\pi\nu/\kappa\) becomes gently time-dependent while retaining the geometric statistics to first order.

\subsection{Even-parity ringdown of Schwarzschild} \label{ssec2.2}
We model the post-merger geometry as a linear perturbation of Schwarzschild driven by the black hole's quasinormal modes (QNMs). Writing
\begin{equation}
    g_{ab}=g^{(0)}_{ab}+\varepsilon \, h_{ab} \quad 0<\varepsilon\ll1, \label{2.11}
\end{equation}
we expand \(h_{ab}\) in tensor spherical harmonics. In this paper, we begin with the even-parity sector, for which the dynamics are encoded in the Zerilli-Moncrief gauge-invariant master function \(\Psi_{\ell m}(t,r)\) \cite{Moncrief:1974am,Martel:2005ir}.

Let \(r_{*}\) be the tortoise coordinate and \(\lambda=(\ell-1)(\ell+2)/2\). The even-parity master field obeys \cite{Zerilli:1970wzz,Martel:2005ir,Berti:2009kk,Konoplya:2011qq}
\begin{equation}
    \left(-\partial_{t}^{2}+\partial_{r_{*}}^{2}-V^{\rm Z}_{\ell}(r)\right)\Psi_{\ell m}(t,r)=0, \label{2.12}
\end{equation}
with the Zerilli potential \cite{Zerilli:1970wzz,Chandrasekhar:1985kt,Kokkotas:1999bd}
\begin{equation}
    V^{\rm Z}_{\ell}(r)=\frac{2\left(1-\frac{2M}{r}\right)}{r^{3}\left(\lambda r+3M\right)^{2}} \left[\lambda^{2}(\lambda+1)r^{3}+3M\lambda^{2}r^{2}+9M^{2}\lambda r+9M^{3}\right]. \label{2.13}
\end{equation}
Ringdown is described by a superposition of homogeneous QNMs with complex frequencies \(\omega_{\ell n}=\omega_{R}-i\omega_{I}\) (\(\omega_{I}>0\)). For a single mode we take \cite{Leaver:1985ax,Nollert:1993zz,Berti:2009kk,Crispino:2007eb}
\begin{equation}
    \Psi_{\ell m}(t,r)=\mathcal{A}_{\ell m} \psi_{\ell}(r) e^{-i\omega t}, \quad \psi_{\ell}\sim \begin{cases} e^{+i\omega r_{*}}, & r\to\infty\ \text{(outgoing)}, \\ e^{-i\omega r_{*}}, & r\to 2M\ \text{(ingoing)},
\end{cases} \label{2.14}
\end{equation}
where \(\mathcal{A}_{\ell m}\) is a (dimensionless) amplitude fixed by the preceding nonlinear dynamics but treated as \(\mathcal{O}(\varepsilon)\) here. The boundary conditions in \eqref{2.14} select the discrete spectrum \(\omega_{\ell n}\).

To display horizon regularity, we use ingoing Eddington-Finkelstein (EF) coordinates (\(v,r,\theta,\phi\)), \(v=t+r_{*}\), so that the background metric reads \cite{Eddington:1924pmh,Finkelstein:1958zz}
\begin{equation}
    ds^{2}= -\left(1-\frac{2M}{r}\right)dv^{2}+2 dv dr+r^{2}d\Omega^{2}. \label{2.15}
\end{equation}
A QNM of the form \(e^{-i\omega t}e^{-i\omega r_{*}}=e^{-i\omega v}\) is manifestly regular on the future horizon \(r=2M\). We therefore reconstruct \(h_{ab}\) from \(\Psi_{\ell m}\) in an EF-regular gauge (equivalent to transforming the standard Regge-Wheeler gauge expressions). All metric amplitudes remain finite at \(r=2M\) for the ingoing solution.

The even-parity perturbation can be written as \cite{Regge:1957td,Zerilli:1970wzz,Martel:2005ir}
\begin{equation}
    h_{ab}=\sum_{\ell m}\left[H_{0}^{\ell m}  Y_{\ell m} (dv)_{a}(dv)_{b} +2H_{1}^{\ell m} Y_{\ell m} (dv)_{(a}(dr)_{b)} +H_{2}^{\ell m} Y_{\ell m} (dr)_{a}(dr)_{b} +K^{\ell m} r^{2} \gamma_{ab}^{\perp} Y_{\ell m} +G^{\ell m} r^{2} Y^{\ell m}_{ab}\right], \label{2.16}
\end{equation}
where \(\gamma_{ab}^{\perp}\) is the metric on \(S^{2}\) and \(Y_{ab}^{\ell m}\) is the even-parity tracefree tensor harmonic. Each coefficient \({H_{0},H_{1},H_{2},K,G}\) is an algebraic combination of \(\Psi_{\ell m}\) and its \(r\)- and \(t\)-derivatives divided by \(\lambda r+3M\). We will only need the contraction of \(h_{ab}\) with outgoing null directions; the explicit (but lengthy) formulas are standard and can be inserted when we evaluate observable coefficients.

The dominant ringdown is the quadrupole. We align the detector on the symmetry axis and take the axisymmetric mode:
\begin{equation}
    \Psi_{20}(t,r)=\mathcal{A}_{20} \psi_{2}(r) e^{-i\omega t},\quad Y_{20}(\theta)=\sqrt{\frac{5}{4\pi}}\frac{1}{2} (3\cos^{2}\theta-1),\quad Y_{20}(0)=\sqrt{\frac{5}{4\pi}}. \label{2.17}
\end{equation}
On the axis (\(\theta=0\)) all azimuthal derivatives vanish, and the only angular dependence is the constant \(Y_{20}(0)\). Consequently, along the axis the nonvanishing components of \(h_{ab}\) reduce to EF scalars multiplying \(Y_{20}(0)\), e.g.
\begin{equation}
    h_{vv}= \mathcal{H}_{vv}(r) Y_{20}(0) e^{-i\omega v},\quad
    h_{vr}= \mathcal{H}_{vr}(r) Y_{20}(0) e^{-i\omega v},\quad
    h_{rr}= \mathcal{H}_{rr}(r) Y_{20}(0) e^{-i\omega v},\ \ldots \label{2.18}
\end{equation}
with \(\mathcal{H}_{ab}(r)\) regular at \(r=2M\) and determined by \(\psi_{2}(r)\).

In the baseline (\ref{ssec2.1}), the detector couples to an outgoing mode with phase \(u=t-r_{*}\). In a time-dependent spacetime, the relevant phase is the solution of the eikonal equation \cite{Wald:1999vt,Carroll_2019}
\begin{equation}
    g^{ab} \partial_{a}u \partial_{b}u=0, \label{2.19}
\end{equation}
normalized to match the usual retarded time at infinity. Linearizing \eqref{2.19} about Schwarzschild with \(g_{ab}=g^{(0)}_{ab}+\varepsilon h_{ab}\) and \(u=u_0+\varepsilon\,\delta u\) gives the transport equation along the background \emph{outgoing}, future-directed null generator
\(k^{a}=-g^{ab}_{(0)}\nabla_{b}u_{0}\):
\begin{equation}
    k^a\nabla_a\delta u=\tfrac12\,h_{ab}\,k^a k^b .
    \label{2.20}
\end{equation}
on $g^{(0)}$. Eq. \eqref{2.20} is the key bridge from ringdown geometry to detector physics: it shows that the correction \(\delta u\) is sourced by the double-null contraction \(h_{ab}k^{a}k^{b}\) of the even-parity metric perturbation.

For the EF background it is convenient to write the background retarded time as
$u_0=v-2r_*(r)$ (with $dr_*/dr=f^{-1}$ and $f(r)=1-2M/r$). The outgoing null generator
associated with the eikonal linearization is then defined geometrically by
\[
k^a=-g_{(0)}^{ab}\nabla_b u_0
=\left(\partial_r\right)^a+\frac{2}{f(r)}\left(\partial_v\right)^a ,
\]
so that $k^a\nabla_a u_0=0$ and $k^a\nabla_a r=1$ (hence $r$ is an affine parameter along
the outgoing rays). The ingoing partner may be chosen as
\[
n^a=-\frac{f(r)}{2}\left(\partial_r\right)^a ,
\]
which obeys $n^a\nabla_a v=0$ and normalizes the dyad as $k\cdot n=-1$. Using \eqref{2.18}, the source term is simply a linear combination of \({\mathcal{H}_{vv} \mathcal{H}_{vr} \mathcal{H}_{rr}}\) evaluated on the axis. Integrating \eqref{2.20} radially from the near-horizon region to the detector sampling radius \(r=r_{c}\) yields
\begin{equation}
    \delta u\Big|_{r_{c},\theta=0} = \Re \!\left[\mathcal{C}_{20}(r_{c})\,e^{-i\omega v_{c}}\right], \quad v_{c}\equiv v(\tau_{c}), \label{2.21}
\end{equation}
where the ringdown response coefficient
\begin{equation}
    \mathcal{C}_{20}(r_{c})=\frac{1}{2}\sqrt{\frac{5}{4\pi}} \mathcal{A}_{20}\int_{2M}^{r_{c}} \mathcal{S}_{20}(r) dr  \label{2.22}
\end{equation}
is determined by the EF-regular combination \(\mathcal{S}_{20}(r)\) of \(\psi_{2}\) and its derivatives (explicit formula supplied where needed). The prefactor $\tfrac12\sqrt{\tfrac{5}{4\pi}}=\tfrac12\,Y_{20}(0)$ is part of the definition of
$\mathcal{C}_{20}$ and is carried consistently throughout (e.g., in the algebraic formula \eqref{4.12} and in the reconstruction and gauge-calibration discussion
of Appendices \ref{apdxB} and \ref{apdxD}). The coefficient \(\mathcal{C}_{20}(r_{c})\) is in general complex (it inherits the QNM phase and the complex normalization \(\mathcal{A}_{20}\)); the physical redshift-map deformation is always the real part of the full harmonic combination \(\mathcal{C}_{20}(r_{c})\,e^{-i\omega v}\), i.e. it is \(\propto e^{-\omega_{I}v}\cos\!\big(\omega_{R}v-\arg\mathcal{C}_{20}(r_{c})\big)\).

\begin{figure}
    \centering
    \includegraphics[width=0.6\linewidth]{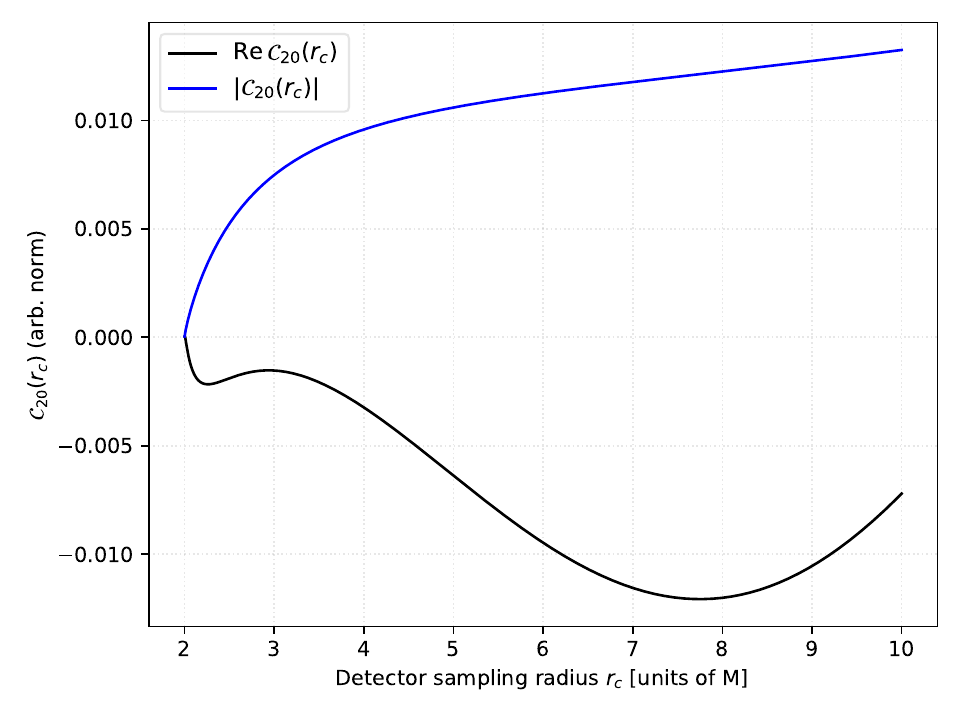}
    \caption{
    Radial response coefficient \(\mathcal{C}_{20}(r_c)\) for the
    fundamental Schwarzschild even-parity QNM \((\ell,m,n)=(2,0,0)\), with
    \(\omega M=0.37367168-0.08896232\,i\) \cite{Berti:2009kk}. Both
    \(\Re\,\mathcal{C}_{20}\) and \(|\mathcal{C}_{20}|\) are shown in
    arbitrary normalization. The coefficient is computed from the
    EF-regular frequency-domain Zerilli solution and the boundary formula
    given in Sec. \ref{ssec4.2}; numerical details and convergence checks
    are collected in Appendix \ref{apdxC20num}.
    }
    \label{fig1}
\end{figure}

The coefficient \(\mathcal{C}_{20}(r_c)\) is the radial response of the
redshift map to the quadrupolar ringdown perturbation. Its magnitude controls
the amplitude with which the curvature perturbation shifts the detector
phase, while its argument fixes the phase of the QNM modulation measured at
the cavity. Figure \ref{fig1} illustrates this response for the fundamental
Schwarzschild even-parity mode \((\ell,m,n)=(2,0,0)\). Close to the horizon,
EF regularity suppresses the response and \(\mathcal{C}_{20}(r_c)\) grows
linearly with \(r_c-2M\). At larger radii the coefficient crosses over toward
the expected radiative falloff. The zero crossings of
\(\mathrm{Re}\,\mathcal{C}_{20}\) reflect the radial phase accumulated by the
ingoing-regular QNM and determine the phase offset of the physical
combination
\[
    \Re\!\left[\mathcal{C}_{20}(r_c)e^{-i\omega v_c}\right].
\]
Thus Fig. \ref{fig1} should be read as a calibration curve for the sampling
radius: it identifies how strongly, and with what phase, a cavity at \(r_c\)
converts the geometric ringdown perturbation into a detector redshift-map
correction.

, the EF-regular radial solution is normalized by
\(\hat\Psi_{20}(2M)=1\) and evaluated with the Fourier convention
\(\Psi_{20}\propto e^{-i\omega v}\). The overall normalization of
\(\mathcal{C}_{20}\) is therefore arbitrary and proportional to the
ringdown amplitude \(\mathcal{A}_{20}\). The plotted curves display only the
radial dependence relevant for the detector response. The closed boundary
formula used to compute \(\mathcal{C}_{20}(r_c)\) is derived in
Sec. \ref{ssec4.2}, while Appendix \ref{apdxC20num} records the numerical
implementation and the checks against the direct transport integral.

With \(u(\tau)=u_{0}(\tau)+\varepsilon \, \delta u(\tau)\), the interaction phase becomes
\begin{equation}
    \Phi(\tau)=\nu u(\tau)\pm \omega_{A}\tau =\nu u_{0}(\tau)\pm \omega_{A}\tau +\varepsilon \, \nu \delta u(\tau)  \label{2.23}
\end{equation}
so that all excitation/absorption amplitudes inherit a parametric modulation through \(\delta u(\tau)\). As we show in Section \ref{sec3}, to first order in \(\varepsilon\), this produces a multiplicative correction to the Boltzmann exponent governing the detailed-balance ratio, oscillating at \(\omega_{R}\) and decaying at \(\omega_{I}\), while preserving the geometric single-mode statistics.

We choose \(\mathcal{A}_{20}\) such that the gauge-invariant energy content in the ringdown slice is \(\propto \varepsilon^{2}|\mathcal{A}_{20}|^{2}\). All observable corrections in this paper scale linearly with \(\varepsilon\mathcal{A}_{20}\) via \(\mathcal{C}_{20}(r_{c})\) in \eqref{2.22}. The static limit \(\varepsilon\to0\) (or late-time limit \(v\to\infty)\) reduces to \ref{ssec2.1} exactly.

\subsection{Detector kinematics and mode geometry} \label{ssec2.3}
We model each probe as a two-level Unruh-DeWitt-type system with energy gap \(\omega_{A}>0\) \cite{DeWitt_1979}, following radial free fall through a small, weakly leaky single-mode cavity centered at radius \(r=r_{c}\) on the symmetry axis (\(\theta=0\)). The coupling to the quantum field is switched on only during the brief transit across the cavity.

Let \(E\) be the conserved specific energy of a radial timelike geodesic in Schwarzschild. The kinematics are \cite{Chandrasekhar:1985kt}
\begin{equation}
    \frac{dt}{d\tau}=\frac{E}{1-2M/r},\quad \frac{dr}{d\tau}=-\sqrt{E^{2}-\left(1-\frac{2M}{r}\right)},\quad \frac{d\phi}{d\tau}=0. \label{2.24}
\end{equation}
We take \(E=1\) (fall from rest at infinity) as the baseline; the \(E\neq 1\) generalization will only rescale subleading prefactors. The tortoise coordinate and null times are as in \eqref{2.2},
\begin{equation}
    r_{*}=r+2M\ln \Big|\frac{r}{2M}-1\Big| \quad u=t-r_{*},\quad v=t+r_{*}. \label{2.25}
\end{equation}
Along the geodesic, near the horizon \(r\to2M\), the retarded time pulled back to the worldline has the universal logarithmic behavior
\begin{equation}
    u(\tau)=u_{0}-\frac{1}{\kappa}\ln \left[\kappa(\tau_{H}-\tau)\right]+\mathcal{O}(\tau_{H}-\tau), \label{2.26}
\end{equation}
where \(\kappa=1/(4M)\) and \(\tau_{H}\) is the finite proper time of horizon crossing. In contrast, the advanced time remains finite,
\begin{equation}
    v(\tau)=v_{H}+\mathcal{O}(\tau_{H}-\tau), \label{2.27}
\end{equation}
so that QNM phases \(\propto e^{-i\omega v}\) are regular on the trajectory at the horizon.

When the ringdown perturbation is present, \(u\) solves the linearized eikonal equation \eqref{2.19}, giving \(u=u_{0}+\varepsilon \, \delta u\). Evaluating \eqref{2.21} on the axis at the cavity crossing time \(\tau=\tau_{c}\) yields
\begin{equation}
    \delta u(\tau_{c})= \Re \!\left[\mathcal{C}_{20}(r_{c})\,e^{-i\omega v_{c}}\right], \quad v_{c}\equiv v(\tau_{c}), \label{2.28}
\end{equation}
where \(\mathcal{C}_{20}(r_{c})\) encodes the integrated even-parity source along the outgoing congruence from \(2M\) to \(r_{c}\) [cf. \eqref{2.22}].

We use an EF-regular null dyad \({k^{a},n^{a}}\) adapted to outgoing/ingoing directions,
\begin{equation}
    k^{a}\nabla_{a}u_{0}=0,\quad k^{a}k_{a}=0,\quad n^{a}\nabla_{a}v=0,\quad n^{a}n_{a}=0,\quad k^{a}n_{a}=-1, \label{2.29}
\end{equation}
with \(k^{a}\) tangent to the background outgoing null geodesics. Eq. \eqref{2.20} shows that the only ringdown datum that enters the detector phase is the double-null contraction
\begin{equation}
    \mathcal{H}_{kk}\equiv h_{ab} k^{a}k^{b}, \label{2.30}
\end{equation}
which, under admissible EF-regular even-parity gauge transformations that keep \(k^{a}\) affinely parametrized, shifts by an affine total derivative along \(k^{a}\),
\(\mathcal{H}_{kk}\mapsto \mathcal{H}_{kk}+\frac{d}{dr}\Xi(r)\), so the induced \(\delta u\) changes only by an endpoint term
(Appendix \ref{apdxD}). Physical detector predictions below therefore depend only on calibration-invariant combinations
(e.g.\ \(\Delta(\delta u)\) and \(\dot{\delta u}\)), once we fix the residual endpoint calibration on the cavity worldtube; in practice we impose the convention \(\Xi(r_{c})=0\) (Appendix \ref{apdxD}), which aligns the worldtube phase with the asymptotic retarded-time phase used to define the frequency label \(\nu\) at \(\mathscr{I}^{+}\). In terms of the Zerilli master function \(\Psi_{20}\), \(\mathcal{H}_{kk}\) is an algebraic combination of \(\Psi_{20}\) and its \(r\)-derivative divided by \(\lambda r+3M\), evaluated at \(\theta=0\); its radial integral builds \(\mathcal{C}_{20}(r_{c})\).

We select a single outgoing mode defined with respect to the retarded time \(u\) at future null infinity. Operationally, we use a narrow wavepacket with central frequency \(\nu>0\) and envelope \(f(u)\) supported during the atom's transit,
\begin{equation}
    \hat{\Phi}_{\nu}(u)=\int \frac{d\tilde{\nu}}{2\pi} \tilde{f}(\tilde{\nu}-\nu) \hat{a}_{\tilde{\nu}} e^{-i\tilde{\nu}u}+\text{H.c.},\quad \int du |f(u)|^{2}=1, \label{2.31}
\end{equation}
\([\hat{a}_{\tilde{\nu}} \hat{a}^{\dagger}_{\tilde{\nu}'}]=2\pi \delta(\tilde{\nu}-\tilde{\nu}')\).
We denote by \(\Delta\nu\) the characteristic bandwidth of the filter, i.e.\ \(\tilde f(\tilde\nu-\nu)\) has support on \(|\tilde\nu-\nu|\lesssim \Delta\nu\), with \(\Delta\nu\ll \nu\).
In the cavity, the spatial profile is approximately constant over the atomic trajectory, so the relevant worldline pullback is entirely through the phase \(u(\tau)\) and the envelope \(f\!\left(u(\tau)\right)\).

Although ringdown breaks exact time-translation symmetry, the frequency label \(\nu\) is fixed operationally by the (weakly leaky) single-mode cavity: the cavity acts as a narrowband filter for the outgoing sector and thereby selects an outgoing wavepacket whose phase is tracked by the EF-regular eikonal \(u\). Concretely, in the static background, on the cavity worldtube \(r=r_c\) one has \(u_0=v-2r_*(r_c)\), so an outgoing Fourier component satisfies
\(e^{-i\tilde\nu u_0}=e^{+i2\tilde\nu r_*(r_c)}\,e^{-i\tilde\nu v}\);
thus the \(\tilde\nu\)-label defined at \(\mathscr{I}^+\) coincides with the Fourier frequency along the cavity worldtube (and with the cavity proper-time frequency up to the usual redshift factor).
In particular, the residual EF-regular endpoint freedom
\(u\to u+\varepsilon\,\Xi(r_c)\) on the worldtube corresponds to a constant rephasing of the selected packet,
\(\hat{\Phi}_\nu\to e^{-i\nu\,\varepsilon\,\Xi(r_c)}\hat{\Phi}_\nu\),
which may be absorbed into the cavity-mode operator and does not affect transition probabilities.
We fix this calibration by the convention \(\Xi(r_c)=0\) (Appendix \ref{apdxD}), which is the same retarded-phase convention used to identify the worldtube mode phase with the asymptotic retarded-time phase that defines \(\nu\) at \(\mathscr{I}^{+}\).
In the perturbed geometry we continue to use the outgoing eikonal \(u=u_0+\varepsilon\,\delta u\) (EF-regular at the horizon), so the selected packet is equivalently characterized by the same \(\tilde\nu\)-label transported along the outgoing congruence.
Within the adiabatic window \eqref{2.34}, the ringdown-induced modulation is quasi-static across a single transit, and any additional frequency drift/mixing in the cavity region is suppressed by the geometric-optics small parameters \(\kappa/\nu\) and \(|\omega|/\nu\), contributing only smooth prefactors beyond the order retained in the detailed-balance ratio.

The atom-field interaction in the interaction picture is \cite{Louko:2006zv}
\begin{equation}
    H_{I}(\tau)=g \chi(\tau) \left(\sigma_{+}e^{+i\omega_{A}\tau}+\sigma_{-}e^{-i\omega_{A}\tau}\right) \hat{\Phi}_{\nu} \left(u(\tau)\right), \label{2.32}
\end{equation}
where \(g\) is the (small) coupling  \(\chi(\tau)\) is a smooth switching function that models the transit through the cavity of proper duration \(\Delta\tau_{c}\), and \(\sigma_{\pm}\) acts on the two-level system. To leading order in \(g\), the excitation/de-excitation amplitudes are
\begin{equation}
    \mathcal{A}_{\mathrm{exc}}=g \int d\tau \chi(\tau) e^{+i\omega_{A}\tau} e^{+i\nu u(\tau)},\quad \mathcal{A}_{\mathrm{abs}}=g \int d\tau \chi(\tau) e^{-i\omega_{A}\tau} e^{-i\nu u(\tau)}. \label{2.33}
\end{equation}
\begin{equation}
    \kappa\ll \nu \lesssim \omega_{A},\quad \kappa\,\Delta\tau_{c} \ll 1,\quad \omega_{I}\Delta v_{c}\ll 1,\quad \omega_{R}\Delta v_{c}\ll 1,
    \quad \Delta v_{c}\equiv \left(\frac{dv}{d\tau}\right)_{\tau_{c}}\Delta\tau_{c},
    \label{2.34}
\end{equation}
which ensures: (i) the near-horizon (Rindler) structure controls the integrals; (ii) the full ringdown factor \(e^{-i\omega v(\tau)}\) (phase and envelope) varies adiabatically across a single transit; and (iii) the wavepacket remains narrow in frequency relative to geometric scales.

Using \(u(\tau)=u_{0}(\tau)+\varepsilon \, \delta u(\tau)\) with \eqref{2.26}-\eqref{2.28}, we expand the phases to first order,
\begin{equation}
    e^{\pm i\nu u(\tau)}=e^{\pm i\nu u_{0}(\tau)}\left[1\pm i \varepsilon \, \nu \delta u(\tau)\right]+\mathcal{O}(\varepsilon^{2}). \label{2.35}
\end{equation}
The \(\varepsilon^{0}\) term reproduces the static Schwarzschild baseline of \ref{ssec2.1}. The \(\mathcal{O}(\varepsilon)\) term injects the decaying oscillation \(e^{-i\omega v_{c}}\) through \(\delta u(\tau)\), yielding a linear, parametric modulation of excitation/absorption probabilities that we compute explicitly in Section \ref{sec3}.

To make contact with the steady-state single-mode statistics, we approximate the transit as a compact support window centered at \(\tau_{c}\),
\begin{equation}
    \chi(\tau)=\chi_{0} w \left(\frac{\tau-\tau_{c}}{\Delta\tau_{c}}\right),\quad \int d\tau \chi^{2}(\tau)=1, \label{2.36}
\end{equation}
with \(w\) a fixed smooth bump. For \(\Delta\tau_{c} \kappa\ll 1\) we may evaluate slowly varying factors at \(\tau_{c}\) and use the near-horizon form \eqref{2.26} inside the integral. This leads to the gamma-function structure in the baseline probabilities and, after including \eqref{2.35}, to their ringdown-modulated counterparts.

\section{Perturbative Framework} \label{sec3}
We now develop the linear response of the detector-mode system to an even-parity ringdown perturbation of Schwarzschild. Our strategy is to (i) fix a null frame that is regular on the future horizon, (ii) compute the first-order correction \(\delta u\) to the outgoing eikonal (u) by integrating the linearized transport equation along the background outgoing congruence, and (iii) pull this corrected redshift map \(u(\tau)=u_{0}(\tau)+\varepsilon \, \delta u(\tau)\) onto the detector worldline. Subsections 3.2-3.3 will use \(u(\tau)\) to evaluate the excitation and absorption amplitudes to \(\mathcal{O}(\varepsilon)\) and derive the modulated detailed-balance relation.

\subsection{Null frame and redshift map} \label{ssec3.1}
In ingoing Eddington-Finkelstein (EF) coordinates (\(v,r,\theta,\phi\)) with background metric \(ds^{2}=-(1-2M/r)dv^{2}+2 dv dr+r^{2}d\Omega^{2}\), we choose the EF-regular null dyad
\begin{equation}
    k^a=-g_{(0)}^{ab}\nabla_b u_0
=\left(\partial_r\right)^a+\frac{2}{f(r)}\left(\partial_v\right)^a ,
\quad
n^a=-\frac{f(r)}{2}\left(\partial_r\right)^a ,
\quad
k^a k_a=n^a n_a=0,\quad k^a n_a=-1,
    \label{3.1}
\end{equation}
The background outgoing eikonal is $u_0=v-2r_*(r)$, so by construction $k^a\nabla_a u_0=0$.
With the normalization $k^a\nabla_a r=1$, the areal radius $r$ is an affine parameter along
the outgoing null rays and
\[
k^a\nabla_a=\frac{d}{dr}=\left(\partial_r\right)_{u_0}
=\partial_r+\frac{2}{f(r)}\,\partial_v . \label{3.2}
\]
Hence
\begin{equation}
    \dfrac{d}{dr}\,\delta u(v,r,\theta)=\frac{1}{2} h_{kk}(v,r,\theta). \label{3.3}
\end{equation}
We integrate from the future horizon \(r=2M\) to the sampling radius \(r=r_{c}\), fixing the integration constant by the EF-regular condition \(\delta u|_{r=2M}=0\):
\begin{equation}
    \delta u(v, r_{c},\theta)=\frac{1}{2}\int_{2M}^{r_{c}}dr h_{kk}(v,r,\theta). \label{3.4}
\end{equation}

For the (\(\ell,m)=(2 0)\) even-parity QNM, the metric perturbation reconstructed in an EF-regular gauge has the schematic form \(h_{ab}\sim \mathcal{H}_{ab}(r) Y_{20}(\theta) e^{-i\omega v}+\text{c.c.}\). On the symmetry axis (\(\theta=0\)), \(Y_{20}(0)=\sqrt{5/4\pi}\) and all angular derivatives vanish, so the only datum that sources \(\delta u\) is the double-null contraction
\begin{equation}
    h_{kk}(v,r,\theta=0)= \sqrt{\frac{5}{4\pi}} \mathcal{S}_{20}(r) \Re \left[e^{-i\omega v}\right], \label{3.5}
\end{equation}
where \(\mathcal{S}_{20}(r)\) is a regular EF combination of the Zerilli master function \(\Psi_{20}\) and its radial derivative divided by \(\lambda r+3M\) (as defined in \ref{ssec2.2}). Substituting \eqref{3.5} into \eqref{3.4} yields
\begin{equation}
    \delta u(v, r_{c},0)= \Re \!\left[\mathcal{C}_{20}(r_{c})\,e^{-i\omega v}\right],\quad
    \mathcal{C}_{20}(r_{c})\equiv \frac{1}{2}\sqrt{\frac{5}{4\pi}}\int_{2M}^{r_{c}} \mathcal{S}_{20}(r) dr. \label{3.6}
\end{equation}
Eq. \eqref{3.6} is the promised redshift map: the perturbation imparts a decaying oscillatory correction to the retarded time with envelope \(e^{-\omega_{I}v}\) and carrier \(\omega_{R}\).

Let \(\tau\mapsto x^{a}(\tau)\) be the detector worldline (radial free fall with \(E=1)\). Pulling back \eqref{3.6} to the detector at the cavity crossing time \(\tau=\tau_{c}\) gives
\begin{equation}
    u(\tau_{c})=u_{0}(\tau_{c})+\varepsilon \, \delta u(\tau_{c}), \quad
    \delta u(\tau_{c})= \Re \!\left[\mathcal{C}_{20}(r_{c})\,e^{-i\omega v_{c}}\right],\quad v_{c}\equiv v(\tau_{c}). \label{3.7}
\end{equation}
Near the horizon, the background map is the universal logarithm
\begin{equation}
    u_{0}(\tau)=u_{*}-\kappa^{-1}\ln \left[\kappa(\tau_{H}-\tau)\right]+\mathcal{O}(\tau_{H}-\tau), \label{3.8}
\end{equation}
so the redshift rate along the worldline splits into baseline plus perturbation,
\begin{equation}
    \frac{du}{d\tau}=\frac{1}{\kappa(\tau_{H}-\tau)}+\varepsilon \ \frac{d}{d\tau}\delta u(\tau)+\mathcal{O}(\varepsilon^{2}). \label{3.9}
\end{equation}
Using \eqref{3.6},
\begin{equation}
    \frac{d}{d\tau}\delta u(\tau)
    =\left[\frac{dr}{d\tau}\,\Re\!\left(\mathcal{C}_{20}'(r)\,e^{-i\omega v}\right)
    +\frac{dv}{d\tau}\,\Re\!\left(-i\omega\,\mathcal{C}_{20}(r)\,e^{-i\omega v}\right)\right]_{\tau}. \label{3.10}
\end{equation}
which stays finite as \(\tau\to\tau_{H}\) because \(v(\tau)\) is regular there. The hierarchy \(\omega_{R}\Delta v_{c}\ll1\) and \(\omega_{I}\Delta v_{c}\ll1\) (equivalently \(|\omega|\Delta v_{c}\ll1\), from \ref{ssec2.3}) ensures that across a single cavity transit the ringdown factor \(e^{-i\omega v(\tau)}\) is quasi-constant, so \(\delta u(\tau)\) varies slowly compared with the near-horizon logarithmic growth of \(u_{0}(\tau)\).

Under even-parity gauge transformations that preserve EF regularity and the normalization of the congruence, the source \(h_{kk}\) shifts by an affine total derivative along \(k^{a}\), so the integrated eikonal correction \(\delta u\) obtained via \eqref{3.4} changes only by an endpoint constant. This constant is precisely the retarded-phase calibration on the cavity worldtube; adopting the operational convention \(\Xi(r_{c})=0\) (equivalently \(\xi_{b}k^{b}\!\mid_{r=r_{c}}=0\)) renders \(\delta u\) and the ensuing detailed-balance modulation invariant within the admissible EF-regular gauge class (Appendix \ref{apdxD}).

\subsection{Outgoing-sector correlators: full vs.\ mode-projected}
\label{ssec3.2}

The detector observable considered here is a \emph{single-mode} response.
The cavity does not sample the full outgoing quantum field; it projects onto
a narrow outgoing wavepacket centered at asymptotic frequency \(\nu\). It is
therefore useful to distinguish the local Hadamard structure of the outgoing
sector from the mode-projected correlator that enters the transition
probabilities.

Near the horizon, the outgoing chiral sector of a massless scalar field has
the standard short-distance form in the eikonal coordinate \(u\),
\begin{equation}
    G^{+}_{\rm out}(u,u')
    =
    -\frac{1}{4\pi}
    \frac{1}{(u-u'-i\epsilon)^2}.
    \label{3.11}
\end{equation}
In the Schwarzschild baseline, using the near-horizon map
\[
    u_0(\tau)
    =
    u_*
    -
    \kappa^{-1}
    \ln\!\left[\kappa(\tau_H-\tau)\right]
    +
    \cdots ,
\]
one obtains the usual stationary pulled-back kernel
\begin{equation}
    G^{+}_{0}(s)
    =
    -\frac{\kappa^2}{16\pi}
    \frac{1}{
    \sinh^2\!\left[
    \frac{\kappa}{2}(s-i\epsilon)
    \right]},
    \qquad
    s\equiv \tau-\tau',
    \label{3.12}
\end{equation}
which satisfies the KMS relation at inverse temperature
\(\beta=2\pi/\kappa\). This full-sector expression is used only as the
local Hadamard and KMS reference. The actual transition probabilities below
are computed with the cavity-projected single-mode correlator.

For the filtered outgoing packet defined in Eq. \eqref{2.31}, the relevant
two-point function is
\begin{equation}
    G^{+}_{\nu}(\tau,\tau')
    =
    \langle 0|
    \hat{\Phi}_{\nu}(u(\tau))
    \hat{\Phi}_{\nu}(u(\tau'))
    |0\rangle
    =
    \int\frac{d\tilde{\nu}}{2\pi}
    |\tilde f(\tilde\nu-\nu)|^2
    e^{-i\tilde\nu[u(\tau)-u(\tau')]} .
    \label{3.13}
\end{equation}
For a narrow wavepacket, \(\Delta\nu\ll\nu\), this may be written as
\begin{equation}
    G^{+}_{\nu}(\tau,\tau')
    =
    \mathfrak{f}\!\left(\Delta u\right)
    e^{-i\nu\Delta u},
    \qquad
    \Delta u\equiv u(\tau)-u(\tau'),
    \label{3.14}
\end{equation}
where
\begin{equation}
    \mathfrak{f}(\Delta u)
    =
    \int\frac{d\varpi}{2\pi}
    |\tilde f(\varpi)|^2e^{-i\varpi\Delta u}
    \label{3.15}
\end{equation}
is a smooth envelope. In the narrowband regime, \(\mathfrak{f}\) varies only
on the scale \(\Delta\nu^{-1}\). It therefore contributes a smooth common
factor to excitation and absorption and cancels from the detailed-balance
ratio at the order retained.

In the ringdown geometry,
\[
    u(\tau)=u_0(\tau)+\varepsilon\,\delta u(\tau),
\]
so
\[
    \Delta u
    =
    \Delta u_0
    +
    \varepsilon\,\Delta(\delta u),
    \qquad
    \Delta(\delta u)
    \equiv
    \delta u(\tau)-\delta u(\tau') .
\]
Expanding Eq. \eqref{3.14} to first order gives
\begin{equation}
    G^{+}_{\nu}(\tau,\tau')
    =
    G^{+}_{\nu,0}(\tau,\tau')
    \left[
    1
    -
    i\varepsilon\nu\,\Delta(\delta u)
    \right]
    +
    \mathcal{O}(\varepsilon^2)
    +
    \mathcal{O}\!\left(
    \varepsilon\,\frac{\Delta\nu}{\nu}
    \right),
    \label{3.16}
\end{equation}
where
\begin{equation}
    G^{+}_{\nu,0}(\tau,\tau')
    =
    \mathfrak{f}(\Delta u_0)
    e^{-i\nu\Delta u_0}.
    \label{3.17}
\end{equation}
Thus the leading ringdown correction to the single-mode correlator is the
antisymmetric eikonal difference \(\Delta(\delta u)\).

Using Eq. \eqref{3.6}, the eikonal perturbation at the cavity is
\begin{equation}
    \delta u(\tau)
    =
    \Re\!\left[
    \mathcal{C}_{20}(r_c)e^{-i\omega v(\tau)}
    \right],
    \label{3.18}
\end{equation}
and hence, across a single transit,
\begin{equation}
    \Delta(\delta u)
    =
    \Re\!\left[
    \mathcal{C}_{20}(r_c)
    \left(
    e^{-i\omega v(\tau)}
    -
    e^{-i\omega v(\tau')}
    \right)
    \right].
    \label{3.19}
\end{equation}
When \(|\omega|\Delta v_c\ll1\), this becomes
\begin{equation}
    \Delta(\delta u)
    =
    s\,\dot{\delta u}(\tau_c)
    +
    \mathcal{O}\!\left(|\omega|^2\Delta v_c^2\right),
    \qquad
    s\equiv\tau-\tau',
    \label{3.20}
\end{equation}
with
\begin{equation}
    \dot{\delta u}(\tau_c)
    =
    \left(\frac{dv}{d\tau}\right)_{\tau_c}
    \Re\!\left[
    -i\omega\,\mathcal{C}_{20}(r_c)e^{-i\omega v_c}
    \right].
    \label{eq_deltau-dot-sec32}
\end{equation}
This is the quantity that enters the probability integrals in
Sec. \ref{ssec3.3}.

A general time-dependent geometry can also perturb the field modes and the
state, producing a correction to the two-point function that is not captured
by the eikonal reparameterization alone. We write this schematically as
\begin{equation}
    G^{+}(\tau,\tau')
    =
    G^{+}_{\rm out}
    \!\left(
    u(\tau),u(\tau')
    \right)
    +
    \varepsilon\,\delta G^{+}_{\rm dyn}(\tau,\tau')
    +
    \mathcal{O}(\varepsilon^2).
    \label{eq_dyn-corr-sec32}
\end{equation}
In the present calculation this dynamical state/mode-shape correction is
subleading. The cavity selects a high-frequency outgoing packet satisfying
\[
    \nu\gg\kappa,
    \qquad
    \nu\gg|\omega|,
    \qquad
    \Delta\nu\ll\nu .
\]
The WKB variation of the filtered mode is therefore slow compared with the
carrier phase, and the associated Bogoliubov mixing is suppressed by
\begin{equation}
    \mathcal{O}\!\left(\varepsilon\frac{\kappa}{\nu}\right)
    +
    \mathcal{O}\!\left(\varepsilon\frac{|\omega|}{\nu}\right).
    \label{eq_wkb-remainder-sec32}
\end{equation}
These contributions are included in the remainder terms of Theorem 1. To
the order kept in the main result, the leading physical effect is therefore
the redshift-map deformation
\[
    u_0(\tau)\longrightarrow u_0(\tau)+\varepsilon\,\delta u(\tau).
\]

The exact KMS relation is broken by ringdown because the correlator is no
longer stationary. However, within one transit the breaking is adiabatic:
it is proportional to \(\varepsilon\mathcal{C}_{20}(r_c)e^{-\omega_Iv_c}\)
and is further controlled by \(|\omega|\Delta v_c\ll1\). Thus the response
can be treated as locally stationary to first order, leading to the
ringdown-modulated detailed-balance relation derived below.

\subsection{Interaction probability integrals} \label{ssec3.3}
We compute excitation/de-excitation probabilities to leading order in the coupling $g$.
Because the cavity selects the single outgoing wavepacket $\hat\Phi_\nu$ [Eq.  \eqref{2.31}],
the response is governed by the \emph{mode-projected} correlator $G^{+}_{\nu}$ in
\ref{ssec3.2}, not by the full-field kernel.
With switching $\chi(\tau)$ localized around $\tau=\tau_{c}$ (transit time $\Delta\tau_{c}$),
the probabilities are
\begin{equation}
    P_{\mathrm{exc}}=g^{2}\!\int d\tau d\tau'\,
    \chi(\tau)\chi(\tau')\,e^{+i\omega_{A}(\tau-\tau')}\,
    G^{+}_{\nu}(\tau,\tau'),
    \label{3.21}
\end{equation}
\begin{equation}
    P_{\mathrm{abs}}=g^{2}\!\int d\tau d\tau'\,
    \chi(\tau)\chi(\tau')\,e^{-i\omega_{A}(\tau-\tau')}\,
    G^{+}_{\nu}(\tau,\tau'),
    \label{3.22}
\end{equation}
with $u(\tau)=u_{0}(\tau)+\varepsilon\,\delta u(\tau)$ and $G^{+}_{\nu}$ given by
\eqref{3.14}--\eqref{3.16}.

Introduce mean and difference times $T=\frac12(\tau+\tau')$, $s=\tau-\tau'$, and the
switching autocorrelation
\[
W(s)\equiv \int dT\,\chi(T{+}s/2)\chi(T{-}s/2).
\]
For a narrow, smooth window ($\kappa\Delta\tau_{c}\ll1$) the $T$-dependence factorizes in the
static background, yielding
\begin{equation}
    P_{\mathrm{exc}}^{(0)}=\mathcal{N} F_{-}(\nu,\kappa,\omega_{A}),\quad
    P_{\mathrm{abs}}^{(0)}=\mathcal{N} F_{+}(\nu,\kappa,\omega_{A}),
    \label{3.23}
\end{equation}
where $u\to u_{0}$ and
\begin{equation}
    F_{\mp}(\nu,\kappa,\omega_{A})
    \equiv
    \int_{-\infty}^{+\infty} ds\,W(s)\,e^{\pm i\omega_{A}s}\,
    G^{+}_{\nu,0}\!\left(T{+}\tfrac{s}{2},T{-}\tfrac{s}{2}\right)\Big|_{T=\tau_{c}}
    \simeq
    \int_{-\infty}^{+\infty} ds\,W(s)\,e^{\pm i\omega_{A}s}\,
    e^{-i\nu\,\Delta u_{0}(s)}.
    \label{3.24}
\end{equation}
Here $\Delta u_{0}(s)\equiv u_{0}(\tau_{c}{+}s/2)-u_{0}(\tau_{c}{-}s/2)$; the dependence on
$\kappa$ enters through the universal near-horizon form of $u_{0}(\tau)$, and the narrowband
envelope in \eqref{3.15} contributes only a smooth common prefactor at this order.

Evaluating \eqref{3.24} in the near-horizon/Rindler window (as detailed in Appendix \ref{apdxA})
gives the Scully single-mode detailed balance,
\begin{equation}
    F_{+}=\frac{e^{2\pi\nu/\kappa}}{e^{2\pi\nu/\kappa}-1}\,\mathcal{A}_{0},\quad
    F_{-}=\frac{1}{e^{2\pi\nu/\kappa}-1}\,\mathcal{A}_{0},
    \label{3.25}
\end{equation}
with a smooth, common prefactor $\mathcal{A}_{0}=\mathcal{A}_{0}(\omega_{A},\chi,\Delta\nu)$
(it cancels in ratios; its explicit form depends on the window and detector parameters, not on $\nu$).
Consequently,
\begin{equation}
    \frac{\Gamma_{\mathrm{abs}}^{(0)}}{\Gamma_{\mathrm{exc}}^{(0)}}
    =\frac{P_{\mathrm{abs}}^{(0)}}{P_{\mathrm{exc}}^{(0)}}
    =e^{2\pi\nu/\kappa},
    \label{3.26}
\end{equation}
reproducing the static detailed-balance result of \ref{ssec2.1}.

Using \eqref{3.14}-\eqref{3.16}, expand the \emph{mode-projected} correlator to first order in $\varepsilon$:
\begin{equation}
    G^{+}_{\nu}(\tau,\tau')
    =
    G^{+}_{\nu,0}(\tau,\tau')
    - i\varepsilon\,\nu\,\Delta(\delta u)\,G^{+}_{\nu,0}(\tau,\tau')
    +\mathcal{O}(\varepsilon^{2}),
    \label{3.27}
\end{equation}
where $\Delta(\delta u)\equiv \delta u(\tau)-\delta u(\tau')$ and
\begin{equation}
    G^{+}_{\nu,0}(\tau,\tau')
    \equiv
    \langle 0|\,\hat\Phi_{\nu}(u_{0}(\tau))\,\hat\Phi_{\nu}(u_{0}(\tau'))\,|0\rangle
    =
    \int\frac{d\tilde\nu}{2\pi}\,|\tilde f(\tilde\nu-\nu)|^{2}\,
    e^{-i\tilde\nu\,[u_{0}(\tau)-u_{0}(\tau')]}
    \simeq e^{-i\nu\,\Delta u_{0}(s)} ,
    \label{3.28}
\end{equation}
with $\Delta u_{0}(s)\equiv u_{0}(\tau)-u_{0}(\tau')$ and $s\equiv\tau-\tau'$.

Only the antisymmetric combination $\Delta(\delta u)$ matters. For a slowly varying $\delta u$
across the window, a first-order Taylor expansion around $T$ yields
\begin{equation}
    \Delta(\delta u)=s\,\dot{\delta u}(T)+\mathcal{O}(s^{3}),
    \quad
    \dot{\delta u}(T)\equiv \frac{d}{d\tau}\delta u(\tau)\Big|_{\tau=T}.
    \label{3.29}
\end{equation}
Substituting \eqref{3.27}-\eqref{3.29} into \eqref{3.21}-\eqref{3.22} and factorizing the $T$-integral gives
\begin{equation}
    P_{\mathrm{exc}}=P_{\mathrm{exc}}^{(0)}+\varepsilon\,\dot{\delta u}_{c}\,\mathcal{K}_{-}
    +\mathcal{O}(\varepsilon^{2}),
    \quad
    P_{\mathrm{abs}}=P_{\mathrm{abs}}^{(0)}-\varepsilon\,\dot{\delta u}_{c}\,\mathcal{K}_{+}
    +\mathcal{O}(\varepsilon^{2}),
    \label{3.30}
\end{equation}
where $\dot{\delta u}_{c}\equiv \dot{\delta u}(T=\tau_{c})$ and the spectral response coefficients are
\begin{equation}
    \mathcal{K}_{\mp}
    \equiv
    \mp i\,g^{2}\nu\int_{-\infty}^{+\infty} ds\, s \,e^{\pm i\omega_{A}s}\,
    G^{+}_{\nu,0}\!\left(T{+}\tfrac{s}{2},T{-}\tfrac{s}{2}\right)\Big|_{T=\tau_{c}}
    \underbrace{\int dT \chi(T{+}s/2)\chi(T{-}s/2)}_{\displaystyle \equiv W(s)}.
    \label{3.31}
\end{equation}
Here \(W(s)\) is the autocorrelation of the switching. Using the near-horizon map \(\Delta u_{0}(s)=\kappa^{-1}\ln \left(1+\frac{s}{s_{*}}\right)+\cdots\) with \(s_{*}\sim \kappa^{-1}\) (cf. \ref{ssec3.2}) and the analyticity of \(G^{+}_{\mathrm{out}}\), the \(s\)-integral is convergent and defines smooth functions \(\mathcal{K}_{\mp}(\nu \kappa,\omega_{A} \chi)\). Two important facts follow:
\begin{itemize}
    \item \(\mathcal{K}_{+}=\mathcal{K}_{-} e^{2\pi\nu/\kappa}\) (a direct consequence of the same contour shift that gives \eqref{3.25});
    \item \(\mathcal{K}_{\mp}\in\mathbb{R}\) for any smooth \(W\) with compact support. In particular, \(\mathcal{K}_{\mp}\) (and hence \(\tilde{\alpha}\)) need not be positive; its sign is controlled by the (windowed) detuning between \(\omega_{A}\) and \(\nu \dot u_{0}\) (cf. Appendix \ref{apdxA}).
\end{itemize}
Thus, \eqref{3.30} can be rewritten as multiplicative corrections to the baseline:
\begin{equation}
    P_{\mathrm{exc}}=P_{\mathrm{exc}}^{(0)}\left[1+\varepsilon \, \dot{\delta u}_{c} \tilde{\alpha}(\nu,\omega_{A} \chi)\right],\quad P_{\mathrm{abs}}=P_{\mathrm{abs}}^{(0)}\left[1-\varepsilon \, \dot{\delta u}_{c} \tilde{\alpha}(\nu,\omega_{A} \chi)\right], \label{3.32}
\end{equation}
with the dimensionless coefficient
\begin{equation}
    \tilde{\alpha}(\nu,\omega_{A} \chi)\equiv \frac{\mathcal{K}_{-}}{P_{\mathrm{exc}}^{(0)}}= \frac{\mathcal{K}_{+}}{P_{\mathrm{abs}}^{(0)}}. \label{3.33}
\end{equation}
(An explicit closed form for \(\tilde{\alpha}\) is given in Appendix \ref{apdxA} for the standard Gaussian window; see also \ref{sec5} for numerics.)

Because the interaction is explicitly time-localized by the switching $\chi(\tau)$, it is convenient to define
a coarse-grained (windowed) transition \emph{rate} by dividing by the effective interaction time
\begin{equation}
    \Delta\tau_{\rm eff}\equiv \int_{-\infty}^{+\infty} d\tau\,\chi(\tau)^{2} \ >0,
    \quad
    \Gamma_{\mathrm{exc}}(\tau_c)\equiv \frac{P_{\mathrm{exc}}}{\Delta\tau_{\rm eff}},
    \quad
    \Gamma_{\mathrm{abs}}(\tau_c)\equiv \frac{P_{\mathrm{abs}}}{\Delta\tau_{\rm eff}}.
    \label{3.33a}
\end{equation}
In the long-time limit where $\chi(\tau)$ approaches a unit-amplitude plateau of duration $T$, one has
$\Delta\tau_{\rm eff}\to T$ and \eqref{3.33a} reduces to the usual constant transition rates.
For fixed switching, the same $\Delta\tau_{\rm eff}$ appears in both channels, so the detailed-balance ratio is
\begin{equation}
    \frac{\Gamma_{\mathrm{abs}}}{\Gamma_{\mathrm{exc}}}
    =\frac{P_{\mathrm{abs}}}{P_{\mathrm{exc}}}.
    \label{3.33b}
\end{equation}
All subsequent rate ratios are understood in this windowed sense.

From \eqref{3.6}-\eqref{3.10},
\begin{equation}
    \delta u(\tau)= \Re\!\left[\mathcal{C}_{20}(r_{c})\,e^{-i\omega v(\tau)}\right],\quad
    \dot{\delta u}_{c}= \ \frac{dv}{d\tau}\Big|_{\tau_{c}} \Re \!\left[-i\omega\,\mathcal{C}_{20}(r_{c})\, e^{-i\omega v_{c}}\right], \label{3.34}
\end{equation}
where \(v_{c}=v(\tau_{c})\) and \((dv/d\tau)_{2M} = 1/2\) is finite at the horizon. Writing \(\mathcal{C}_{20}(r_{c})=|\mathcal{C}_{20}(r_{c})|e^{i\delta_{20}(r_{c})}\) with \(\delta_{20}(r_{c})\equiv \arg\mathcal{C}_{20}(r_{c})\), we obtain
\begin{equation}
    \dot{\delta u}_{c}= |\mathcal{C}_{20}(r_{c})| \left(\frac{dv}{d\tau}\right)_{\tau_{c}} e^{-\omega_{I} v_{c}} \, \mathcal{S}\!\left(\omega_{R}v_{c}-\delta_{20}(r_{c})\right). \label{3.35}
\end{equation}
Combining \eqref{3.32}-\eqref{3.35} yields
\begin{equation}
    P_{\mathrm{exc}}=P_{\mathrm{exc}}^{(0)} \left[1+\varepsilon \, \tilde{\alpha}\, |\mathcal{C}_{20}(r_{c})| \left(\frac{dv}{d\tau}\right)_{\tau_{c}} e^{-\omega_{I} v_{c}}  \mathcal{S}\!\left(\omega_{R}v_{c}-\delta_{20}(r_{c})\right)\right], \label{3.36}
\end{equation}
\begin{equation}
    P_{\mathrm{abs}}=P_{\mathrm{abs}}^{(0)} \left[1-\varepsilon \, \tilde{\alpha}\, |\mathcal{C}_{20}(r_{c})| \left(\frac{dv}{d\tau}\right)_{\tau_{c}} e^{-\omega_{I} v_{c}}  \mathcal{S}\!\left(\omega_{R}v_{c}-\delta_{20}(r_{c})\right)\right], \label{3.37}
\end{equation}
with the ringdown shape function
\begin{equation}
    \mathcal{S}(\theta)\equiv -\omega_{I}\cos\theta-\omega_{R}\sin\theta. \label{3.38}
\end{equation}

\begin{figure}
    \centering
    \includegraphics[width=0.6\linewidth]{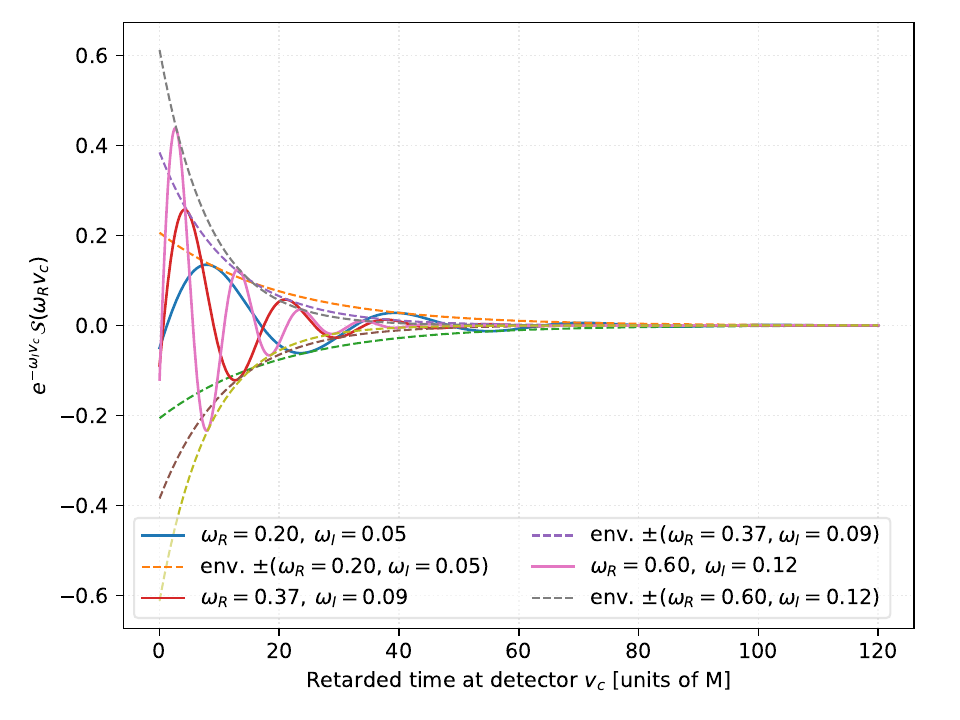}
    \caption{Ringdown kernel \( M(v_c)=e^{-\omega_I v_c}\,\mathcal{S}\!\left(\omega_R v_c-\delta_{20}(r_c)\right) \) for several \((\omega_R,\omega_I)\), with envelopes \( \pm \sqrt{\omega_R^2+\omega_I^2}\,e^{-\omega_I v_c} \) indicating the peak decay. (The constant offset \(\delta_{20}(r_c)\) rigidly shifts the oscillations in \(v_c\).}
    \label{fig2}
\end{figure}

In Fig. \ref{fig2}, changing \(\omega_{R}\) slides the carrier phase inside the universal derivative-shape \(\mathcal{S}\), while \(\delta_{20}(r_{c})=\arg\mathcal{C}_{20}(r_{c})\) provides the overall phase offset set by the EF-regular reconstruction at the cavity radius; the decay rate and hence the visibility window are controlled exclusively by \(\omega_{I}\). The dashed envelopes \(\pm\sqrt{\omega_R^2+\omega_I^2}\,e^{-\omega_I v_c}\) isolate this pure QNM kinematics: geometry and detector specifics enter only through the overall prefactor \(\varepsilon\,\tilde{\alpha}(\nu,\omega_A,\chi)\,\mathcal{C}_{20}(r_c)\,(dv/d\tau)\). In other words, once \(r_c\) (hence \(\mathcal{C}_{20})\) is fixed, varying \(\omega_R\) tunes the timing of peaks while \(\omega_I\) sets the lifetime of the modulation; this is the same separation of roles familiar from the classical ringdown waveform but now imprinted directly on the KMS/detailed-balance exponent.

Taking the ratio of \eqref{3.37} and \eqref{3.36} we find, to \(\mathcal{O}(\varepsilon)\),
\begin{equation}
    \frac{\Gamma_{\mathrm{abs}}}{\Gamma_{\mathrm{exc}}}
    =
    \frac{P_{\mathrm{abs}}}{P_{\mathrm{exc}}}
    =
    e^{2\pi\nu/\kappa}
    \left[
    1
    -
    2\varepsilon \,
    \tilde{\alpha}
    |\mathcal{C}_{20}(r_{c})|
    \left(\frac{dv}{d\tau}\right)_{\tau_{c}}
    e^{-\omega_{I} v_{c}}
    \mathcal{S}\!\left(\omega_{R}v_{c}-\delta_{20}(r_c)\right)
    \right],
    \label{3.39}
\end{equation}
or, equivalently, as an additive modulation of the exponent,
\begin{equation}
    \ln \frac{\Gamma_{\mathrm{abs}}}{\Gamma_{\mathrm{exc}}}
    =
    \frac{2\pi\nu}{\kappa}
    -
    2\varepsilon \,
    \tilde{\alpha}
    |\mathcal{C}_{20}(r_{c})|
    \left(\frac{dv}{d\tau}\right)_{\tau_{c}}
    e^{-\omega_{I} v_{c}}
    \mathcal{S}\!\left(\omega_{R}v_{c}-\delta_{20}(r_c)\right)
    +
    \mathcal{O}(\varepsilon^{2}).
    \label{3.40}
\end{equation}
Thus, the static Boltzmann exponent \(2\pi\nu/\kappa\) is modulated at the ringdown frequency and decays on the QNM timescale. The common prefactor \(\tilde{\alpha}(\nu,\omega_{A} \chi)\) encodes only detector/cavity details; the geometric content resides in \(\mathcal{C}_{20}(r_{c})\) and \((\omega_{R},\omega_{I})\).

\begin{figure}
    \centering
    \includegraphics[width=0.6\linewidth]{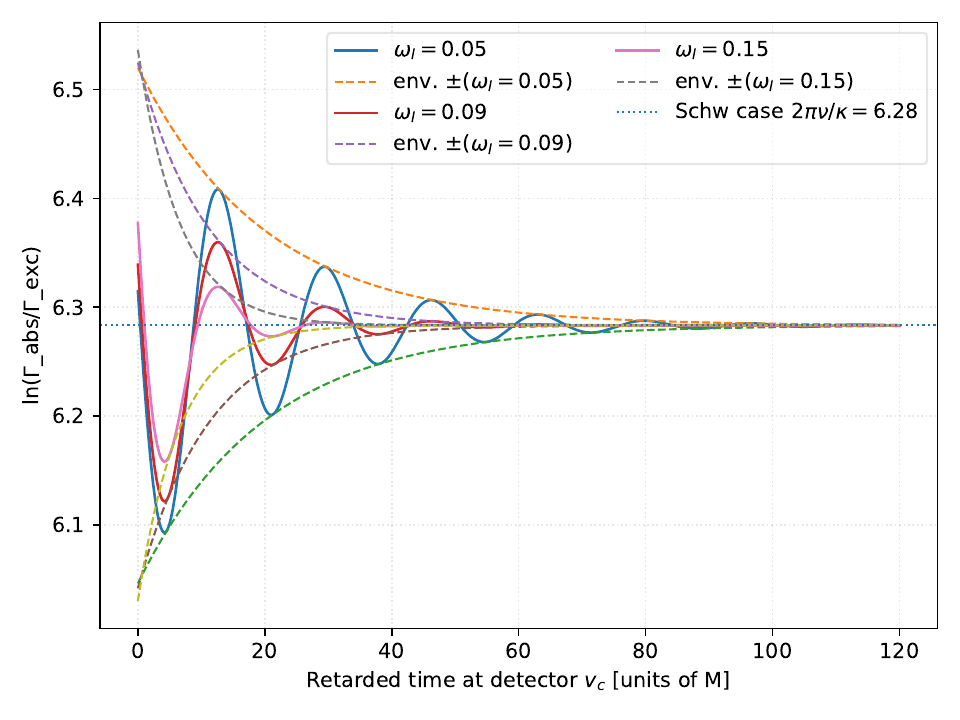}
    \caption{Log detailed-balance ratio \( \ln(\Gamma_{\rm abs}/\Gamma_{\rm exc}) \) for several \(\omega_I\) at fixed \(\omega_R\), with analytic envelopes \( \text{baseline} \pm (2\pi\nu/\kappa)|2\varepsilon\tilde{\alpha}\mathcal{C}_{20}(dv/d\tau)|\sqrt{\omega_R^2+\omega_I^2}\,e^{-\omega_I v_c} \) shown as dashed curves.}
    \label{fig3}
\end{figure}

In Fig. \ref{fig3}, the plot of Eq. \eqref{3.40} makes the physics of damping especially clear: larger \(\omega_I\) shortens the e-folding time of the modulation \(e^{-\omega_I v_c}\) and thus narrows the window over which deviations from the thermal value \(2\pi\nu/\kappa\) can be resolved. Equivalently, the quality factor \(Q\simeq \omega_R/(2\omega_I)\) sets the number of visible oscillations of the detailed-balance exponent before it relaxes back to the stationary Schwarzschild value. The analytic envelopes in the figure summarize the competing scalings: peak height \(\propto \left|\varepsilon\tilde{\alpha}\mathcal{C}_{20}(r_c)(dv/d\tau)\right|\sqrt{\omega_R^2+\omega_I^2}\) and decay constant exactly \(\omega_I\). Operationally, this means detectability can be traded among three knobs: geometry \((\omega_I\), fixed by the background), sampling location \((\mathcal{C}_{20}(r_c))\), and detector windowing \((\tilde{\alpha}\) and \(dv/d\tau)\).

As a final remark, the result hinges on the antisymmetric difference \(\Delta(\delta u)\), not on a constant shift of \(u\); hence, the time derivative \(\dot{\delta u}\) appears. For broader windows the \(T\)-factorization remains valid provided \(|\omega|\,\Delta v_{c}\ll1\) (equivalently \(\omega_{R}\Delta v_{c}\ll1\) and \(\omega_{I}\Delta v_{c}\ll1\)); otherwise Appendix \ref{apdxB} gives the next adiabatic corrections \(\mathcal{O}(|\omega|\Delta v_{c})\). In the late-time limit \(v_{c}\to\infty\), the modulation vanishes and \eqref{3.26} is recovered exactly.

\section{Main Result} \label{sec4}
We now assemble the geometric ingredients of Section \ref{sec3} into a closed
analytic expression for the detector's detailed-balance ratio during
ringdown. The word \textit{universal} will be used here only in a restricted
sense. The result is universal within the specified operational protocol:
\(\ell=2\), \(m=0\), a single outgoing cavity-filtered mode, a narrow
adiabatic transit window, an EF-regular reconstruction, and the endpoint
calibration \(\Xi(r_c)=0\). Within this class, the time dependence is fixed
by the QNM pair \((\omega_R,\omega_I)\), while the radial amplitude and phase
are encoded in the calibrated response coefficient
\(\mathcal{C}_{20}(r_c)\). The overall size and sign of the observable
modulation, however, remain protocol-dependent through the detector gap,
switching function, cavity filter, sampling radius, and worldtube phase
calibration.

\subsection{Ringdown-modulated detailed balance (Theorem 1)}
\label{ssec4.1}

We now state the main result in a form that keeps the operational
assumptions explicit. The observable is the windowed absorption-to-excitation
ratio of a freely falling two-level detector coupled to a cavity-filtered
outgoing mode. It should therefore be interpreted as a detector/cavity
detailed-balance diagnostic, not as a direct statement about the global
Hawking flux.

We assume: (i) a linear even-parity, axisymmetric quadrupolar perturbation of
Schwarzschild,
\[
    g_{ab}=g^{(0)}_{ab}+\varepsilon h_{ab},
    \qquad
    0<\varepsilon\ll1,
\]
with QNM frequency
\[
    \omega=\omega_R-i\omega_I,
    \qquad
    \omega_I>0;
\]
(ii) an EF-regular reconstruction and the outgoing null generator
\(k^a\) normalized by \(k^a\nabla_a r=1\); (iii) a single outgoing
cavity-filtered wavepacket of central asymptotic frequency \(\nu\), with
\(\Delta\nu\ll\nu\); (iv) a detector crossing a narrow cavity centered at
\(r=r_c\) at advanced time \(v_c\); and (v) the adiabatic and
geometric-optics hierarchy
\begin{equation}
    \kappa\Delta\tau_c\ll1,
    \qquad
    |\omega|\Delta v_c\ll1,
    \qquad
    \nu\gg\kappa,
    \qquad
    \nu\gg|\omega|,
    \qquad
    \Delta v_c
    =
    \left(\frac{dv}{d\tau}\right)_{\tau_c}
    \Delta\tau_c .
    \label{4.0}
\end{equation}
The endpoint phase of the cavity mode is calibrated by the convention
\(\Xi(r_c)=0\), as discussed in Appendix \ref{apdxD}. This fixes the
residual worldtube phase convention used to compare the cavity phase with
the asymptotic retarded-time phase defining \(\nu\).

The EF-regular redshift map at the detector is
\begin{equation}
    u(\tau_c)
    =
    u_0(\tau_c)
    +
    \varepsilon\,\delta u(\tau_c),
    \qquad
    \delta u(\tau_c)
    =
    \Re\!\left[
    \mathcal{C}_{20}(r_c)e^{-i\omega v_c}
    \right],
    \label{4.1}
\end{equation}
where
\begin{equation}
    \mathcal{C}_{20}(r_c)
    =
    \frac12\sqrt{\frac{5}{4\pi}}
    \int_{2M}^{r_c}\mathcal{S}_{20}(r)\,dr,
    \qquad
    \delta_{20}(r_c)
    \equiv
    \arg\mathcal{C}_{20}(r_c).
    \label{4.1a}
\end{equation}
Here \(\mathcal{S}_{20}(r)\) denotes the EF-regular radial source obtained
from the double-null contraction \(h_{ab}k^ak^b\) on the symmetry axis. A
closed boundary expression for \(\mathcal{C}_{20}(r_c)\) is given in
Sec. \ref{ssec4.2}.

The derivative of the eikonal correction along the detector worldline is
\begin{equation}
    \dot{\delta u}(\tau_c)
    =
    \left(\frac{dv}{d\tau}\right)_{\tau_c}
    |\mathcal{C}_{20}(r_c)|
    e^{-\omega_I v_c}
    \mathcal{S}
    \!\left(
    \omega_R v_c-\delta_{20}(r_c)
    \right),
    \label{4.2}
\end{equation}
with
\begin{equation}
    \mathcal{S}(\theta)
    =
    -\omega_I\cos\theta
    -
    \omega_R\sin\theta .
    \label{4.2a}
\end{equation}
Let \(\tilde{\alpha}=\tilde{\alpha}(\nu,\omega_A,\chi)\) be the
dimensionless, possibly signed, switching-dependent single-mode response
coefficient defined in Appendix \ref{apdxA}. It contains the detector gap,
switching function, and wavepacket details, but not the ringdown geometry.

Then, to first order in \(\varepsilon\), the windowed rate ratio satisfies
\begin{equation}
    \frac{\Gamma_{\rm abs}}{\Gamma_{\rm exc}}
    =
    \exp\!\left(\frac{2\pi\nu}{\kappa}\right)
    \left[
    1
    -
    2\varepsilon\,
    \tilde{\alpha}\,
    |\mathcal{C}_{20}(r_c)|
    \left(\frac{dv}{d\tau}\right)_{\tau_c}
    e^{-\omega_I v_c}
    \mathcal{S}
    \!\left(
    \omega_R v_c-\delta_{20}(r_c)
    \right)
    +
    \mathcal{E}
    \right],
    \label{4.3}
\end{equation}
or, equivalently,
\begin{equation}
    \ln
    \frac{\Gamma_{\rm abs}}{\Gamma_{\rm exc}}
    =
    \frac{2\pi\nu}{\kappa}
    -
    2\varepsilon\,
    \tilde{\alpha}\,
    |\mathcal{C}_{20}(r_c)|
    \left(\frac{dv}{d\tau}\right)_{\tau_c}
    e^{-\omega_I v_c}
    \mathcal{S}
    \!\left(
    \omega_R v_c-\delta_{20}(r_c)
    \right)
    +
    \mathcal{R}.
    \label{4.4}
\end{equation}
The controlled remainders are
\begin{align}
    \mathcal{E},\mathcal{R}
    =
    &\;
    \mathcal{O}(\varepsilon^2)
    +
    \mathcal{O}(\varepsilon|\omega|\Delta v_c)
    +
    \mathcal{O}\!\left(\varepsilon(\kappa\Delta\tau_c)^2\right)
    \nonumber\\
    &+
    \mathcal{O}\!\left(\varepsilon\frac{\kappa}{\nu}\right)
    +
    \mathcal{O}\!\left(\varepsilon\frac{|\omega|}{\nu}\right)
    +
    \mathcal{O}\!\left(\varepsilon\frac{\Delta\nu}{\nu}\right).
    \label{4.5}
\end{align}

The leading coefficient multiplying the damped sinusoid in Eq. \eqref{4.4}
has a simple physical interpretation. It factorizes into four pieces:
\[
    -2\varepsilon\,
    \tilde{\alpha}(\nu,\omega_A,\chi)\,
    |\mathcal{C}_{20}(r_c)|
    \left(\frac{dv}{d\tau}\right)_{\tau_c}.
\]
The factor \(\varepsilon\) is the linear ringdown amplitude. The coefficient
\(\tilde{\alpha}\) is the detector/cavity transduction factor: it tells how
efficiently a small change in the eikonal phase is converted into a change
of the absorption-to-excitation ratio for the chosen gap, switching function,
and wavepacket. The factor \(|\mathcal{C}_{20}(r_c)|\) is the geometric
redshift-map susceptibility of the sampling radius to the quadrupolar
perturbation. Finally, \((dv/d\tau)_{\tau_c}\) converts the QNM variation in
advanced time into the detector's proper-time response. Thus the QNM
frequency and damping determine the temporal template, whereas the leading
coefficient determines how strongly the chosen operational apparatus reads
out that geometric template.

The proof follows directly from the ingredients developed in
Sec. \ref{sec3}. The mode phase is expanded as
\[
    e^{\pm i\nu u(\tau)}
    =
    e^{\pm i\nu u_0(\tau)}
    \left[
    1
    \pm
    i\varepsilon\nu\,\delta u(\tau)
    \right]
    +
    \mathcal{O}(\varepsilon^2).
\]
Only the antisymmetric difference
\(\delta u(\tau)-\delta u(\tau')\) contributes to the first-order correction
to the probability ratio. In the adiabatic window this difference is
\(s\,\dot{\delta u}(\tau_c)\) to leading order, with
\(\dot{\delta u}(\tau_c)\) given by Eq. \eqref{4.2}. The remaining
windowed single-mode integrals define the common coefficient
\(\tilde{\alpha}\). Since the same effective interaction time divides both
probabilities, the windowed rate ratio equals the probability ratio.

Equation \eqref{4.4} is the central result. It says that the static
Schwarzschild detailed-balance exponent \(2\pi\nu/\kappa\) remains the
baseline, while ringdown adds a small QNM-template correction. The correction
oscillates at \(\omega_R\), decays at \(\omega_I\), and has a calibrated
radial amplitude and phase determined by \(\mathcal{C}_{20}(r_c)\).

Several immediate consequences follow.

First, when \(\varepsilon\to0\) or \(v_c\to\infty\),
\begin{equation}
    \frac{\Gamma_{\rm abs}}{\Gamma_{\rm exc}}
    \longrightarrow
    \exp\!\left(\frac{2\pi\nu}{\kappa}\right),
    \label{4.6}
\end{equation}
recovering the static detector/cavity detailed-balance relation.

Second, detector details enter only through
\(\tilde{\alpha}(\nu,\omega_A,\chi)\). The exponent itself is controlled by
the selected outgoing mode frequency \(\nu\), as in the static
single-mode baseline.

Third, repeated detector passages through a weakly leaky cavity still yield a
geometric photon distribution, but with a time-dependent parameter,
\begin{equation}
    p_n(v_c)
    =
    \left[
    1-e^{-2\xi(v_c)}
    \right]
    e^{-2\xi(v_c)n},
    \qquad
    2\xi(v_c)
    =
    \ln
    \frac{\Gamma_{\rm abs}}{\Gamma_{\rm exc}} .
    \label{4.7}
\end{equation}
Thus the occupation statistics inherit the same damped QNM modulation.

Finally, this modulation is distinguishable from a generic slow instrumental
drift only when the QNM template and radial calibration are used. A true
ringdown contribution has the coherent form
\begin{equation}
    \delta[2\xi]_{\rm RD}(v_c)
    =
    -2\varepsilon\,
    \tilde{\alpha}\,
    |\mathcal{C}_{20}(r_c)|
    \left(\frac{dv}{d\tau}\right)_{\tau_c}
    e^{-\omega_I v_c}
    \mathcal{S}
    \!\left(
    \omega_R v_c-\delta_{20}(r_c)
    \right),
    \label{4.8}
\end{equation}
with the same \((\omega_R,\omega_I)\) across calibrated sampling radii. A
single uncalibrated channel cannot by itself exclude an artificial detuning
drift chosen to mimic the same damped sinusoid. Operational identification
therefore requires either independent monitoring of the cavity frequency or
multi-radius consistency of the amplitude and phase through
\(\mathcal{C}_{20}(r_c)\).

\subsection{Coefficient \texorpdfstring{\(\mathcal{C}_{20}(r_{c})\)}{}: closed-form expression} \label{ssec4.2}
Recall from \eqref{4.1} that along the symmetry axis,
\begin{equation}
    \delta u(\tau_{c})=\Re\!\left[\mathcal{C}_{20}(r_{c})\,e^{-i\omega v_{c}}\right],\quad
    \mathcal{C}_{20}(r_{c})=\frac{1}{2}\sqrt{\frac{5}{4\pi}}\int_{2M}^{r_{c}}  \mathcal{S}_{20}(r)\, dr.  \label{4.9}
\end{equation}
where \(\mathcal{S}_{20}(r)\) is the EF-regular double-null contraction \(h_{kk}\equiv h_{ab}k^{a}k^{b}\) divided by the factor \(Y_{20}(0) \Re[e^{-i\omega v}]\) (cf. \ref{ssec3.1}). In this subsection, we eliminate the integral and give \(\mathcal{C}_{20}(r_{c})\) algebraically in terms of the Zerilli master function and its radial derivative at \(r=r_{c}\).

For even-parity vacuum perturbations, the EF-regular metric reconstruction from the frequency-domain Zerilli-Moncrief function \(\Psi_{20}(r) e^{-i\omega v}\) implies that the axis data entering the eikonal transport equation,
\begin{equation}
    k^{a}\nabla_{a}\delta u=\frac{1}{2}h_{kk}, \label{4.10}
\end{equation}
can be written as a radial total derivative built from \(\Psi_{20}\) and \(\partial_{r}\Psi_{20}\):
\begin{equation}
    h_{kk}(v,r,\theta=0)=\Re\!\left\{\dfrac{d}{dr}\Big[a_2(r)\Psi_{20}(r)+b_2(r)\partial_r\Psi_{20}(r)\Big]\,e^{-i\omega v}\right\}. \label{4.11}
\end{equation}
Here \(a_{2}(r)\) and \(b_{2}(r)\) are rational functions of \(r\) and \(M\) that are regular at \(r=2M\); for \(\ell=2\) they depend on the usual \(\lambda=(\ell-1)(\ell+2)/2=2\) only through the combination \(\Lambda(r)\equiv  \lambda r+3M = 2r+3M\). They result from an algebraic match of the EF-gauge reconstruction to the Moncrief gauge invariant (derivation in Appendix \ref{apdxB}). Eq. \eqref{4.11} is the key simplification: since $k^a\nabla_a r=1$, we may use $r$ as affine parameter
along the outgoing rays, so $k^a\nabla_a=d/dr$ and the transport equation integrates directly to a boundary term.

Substituting \eqref{4.11} into \eqref{4.9} and using EF regularity of \(\Psi_{20}\) at the future horizon (ingoing condition) we obtain
\begin{equation}
    \mathcal{C}_{20}(r_{c})=\frac{1}{2}\sqrt{\frac{5}{4\pi}} \left[a_{2}(r_{c}) \Psi_{20}(r_{c})+b_{2}(r_{c}) \partial_{r}\Psi_{20}(r_{c})\right]. \label{4.12}
\end{equation}
The contribution from \(r=2M\) vanishes because the ingoing EF solution is finite there and \(a_{2}(r) ,b_{2}(r)\) are regular. Writing \(\Psi_{20}(t,r)=\mathcal{A}_{20} \psi_{2}(r) e^{-i\omega t}\) with the QNM normalization of \ref{ssec2.2}  \eqref{4.12} factorizes as
\begin{equation}
    \mathcal{C}_{20}(r_{c})=\frac{1}{2}\sqrt{\frac{5}{4\pi}} \mathcal{A}_{20} \left[a_{2}(r_{c}) \psi_{2}(r_{c})+b_{2}(r_{c}) \psi_{2}'(r_{c})\right]. \label{4.13}
\end{equation}

For completeness as well as to enable direct checks and numerics, we record the explicit rational functions \(a_{2}(r)\) and \(b_{2}(r)\) in terms of (M) (details in Appendix \ref{apdxB}):
\begin{equation}
    a_{2}(r)=\frac{(r-2M)\left(2r^{2}+6Mr+6M^{2}\right)}{r^{2} \Lambda(r)^{2}},\quad b_{2}(r)=\frac{(r-2M)^{2}}{r \Lambda(r)^{2}},\quad \Lambda(r)=2r+3M. \label{4.14}
\end{equation}
With these, \eqref{4.12}-\eqref{4.13} give a fully algebraic \(\mathcal{C}_{20}(r_{c})\) that is manifestly regular at the horizon and straightforward to evaluate at any sampling radius \(r_{c}>2M\).

The pair \((a_{2},b_{2})\) is unique (up to addition of terms that vanish by the Zerilli equation) under the requirements: (i) EF regularity; (ii) dependence only on \(\Psi_{20}\) and \(\partial_{r}\Psi_{20}\); (iii) correct transformation under rescalings of the master field; and (iv) reproduction of the RW-Zerilli reconstruction in the static limit \(\omega\to0\) (details in Appendix \ref{apdxB}).

Two limits are instructive: For \(r_{c}=2M+\delta r\) with \(\delta r\ll 2M\),
\begin{equation}
    \mathcal{C}_{20}(r_{c}) =\frac{1}{2}\sqrt{\frac{5}{4\pi}} \mathcal{A}_{20} a_{2}'(2M) \psi_{2}(2M)(r_{c}-2M)+\mathcal{O} \left((r_{c}-2M)^{2}\right), \label{4.15}
\end{equation}
since \(b_{2}(r)=\mathcal{O}\left((r-2M)^{2}\right)\). Thus \(\mathcal{C}_{20}\) grows linearly with the cavity altitude above the horizon, as required by regularity. For \(r_{c}\gg M\), using the asymptotic QNM behavior \(\psi_{2}(r)\sim e^{+i\omega r_{*}}\) and \(\psi_{2}'\sim i\omega \psi_{2}\), \eqref{4.14} gives
\begin{equation}
    \mathcal{C}_{20}(r_{c})=\frac{1}{2}\sqrt{\frac{5}{4\pi}} \mathcal{A}_{20} \frac{\hat{A}+i\hat{B}\omega r_{c}}{\lambda^{2} r_{c}}\psi_{2}(r_{c})+\mathcal{O} \left(\frac{M}{r_{c}^{2}}\right), \label{4.16}
\end{equation}
so \(\mathcal{C}_{20}\) decays as \(r_{c}^{-1}\) (modulo the oscillatory factor in \(\psi_{2})\).

Given a choice of QNM normalization \(\mathcal{A}_{20}\) and the radial solution \(\psi_{2}(r)\) of the Zerilli equation with ingoing-horizon/outgoing-infinity boundary conditions, \eqref{4.13}-\eqref{4.14} provide a single-line evaluation of \(\mathcal{C}_{20}(r_{c})\). No line integral is needed; only \(\psi_{2}\) and \(\psi_{2}'\) at \(r_{c}\) enter. In numerical work (\ref{sec5}), we compute these via standard Frobenius-Leaver series or direct frequency-domain integration and confirm the near-horizon linear rise \eqref{4.15} and the far-zone decay \eqref{4.16}.

\subsection{Static-limit check}
\label{ssec4.3}

We now verify that the modulation in Eq. \eqref{4.4} disappears in the
limits in which no genuinely time-dependent quadrupolar ringdown remains.
This check is important because the first term in Eq. \eqref{4.4} is the
static detector/cavity detailed-balance exponent, while the second term is
the time-dependent QNM correction.

There are three relevant limits.

First, in the zero-amplitude limit,
\[
    \varepsilon\to0,
\]
the correction in Eq. \eqref{4.4} vanishes linearly in \(\varepsilon\), and
one recovers
\begin{equation}
    \frac{\Gamma_{\rm abs}}{\Gamma_{\rm exc}}
    =
    \exp\!\left(\frac{2\pi\nu}{\kappa}\right)
    +
    \mathcal{O}(\varepsilon^2)
    +
    \mathcal{R}.
    \label{4.18}
\end{equation}

Second, in the late-time limit at fixed perturbation amplitude,
\[
    v_c\to\infty,
\]
the QNM factor decays as \(e^{-\omega_Iv_c}\), with \(\omega_I>0\). Hence
the ringdown contribution again disappears and the same static relation is
obtained.

Third, consider a stationary quadrupolar perturbation. Formally taking
\(\omega\to0\) in the quadrupolar sector makes the eikonal correction
time-independent:
\begin{equation}
    \delta u(\tau)
    =
    \Re\!\left[
    \mathcal{C}_{20}(r_c)
    \right]
    +
    \mathcal{O}(\omega).
    \label{4.19}
\end{equation}
The first-order correction to the single-mode probabilities depends on the
antisymmetric difference
\begin{equation}
    \Delta(\delta u)
    =
    \delta u(\tau)-\delta u(\tau'),
    \label{4.20}
\end{equation}
which vanishes for a constant \(\delta u\). Equivalently, a constant shift of
\(u\) produces only a constant phase redefinition of the selected mode, and
this phase cancels from transition probabilities and from the detailed-
balance ratio. Therefore a strictly stationary quadrupolar perturbation does
not produce the QNM modulation in Eq. \eqref{4.4}.

These limits show that the second term in Eq. \eqref{4.4} is not a
renormalization of the static Schwarzschild temperature or of the
stationary detailed-balance baseline. It is a genuinely time-dependent
ringdown correction to the detector/cavity observable.

A separate comment is needed for static monopole perturbations. The
quadrupolar sector studied here does not change the Schwarzschild mass at
linear order. A true static mass perturbation belongs instead to the
monopole sector and changes the stationary background itself. If
\[
    M_{\rm eff}
    =
    M+\varepsilon\,\delta M,
\]
then
\begin{equation}
    \kappa_{\rm eff}
    =
    \frac{1}{4M_{\rm eff}}
    =
    \kappa
    \left(
    1-\varepsilon\frac{\delta M}{M}
    \right)
    +
    \mathcal{O}(\varepsilon^2),
    \label{4.21}
\end{equation}
and the stationary detailed-balance exponent becomes
\begin{equation}
    \frac{2\pi\nu}{\kappa_{\rm eff}}
    =
    \frac{2\pi\nu}{\kappa}
    \left(
    1+\varepsilon\frac{\delta M}{M}
    \right)
    +
    \mathcal{O}(\varepsilon^2).
    \label{4.22}
\end{equation}
Such a monopole shift changes the static baseline; it is not the
decaying-oscillatory quadrupolar signal derived in this work. If both
effects are present, the appropriate separation is
\begin{equation}
    \ln
    \frac{\Gamma_{\rm abs}}{\Gamma_{\rm exc}}
    =
    \frac{2\pi\nu}{\kappa_{\rm eff}}
    -
    2\varepsilon\,
    \tilde{\alpha}\,
    |\mathcal{C}_{20}(r_c)|
    \left(\frac{dv}{d\tau}\right)_{\tau_c}
    e^{-\omega_I v_c}
    \mathcal{S}
    \!\left(
    \omega_Rv_c-\delta_{20}(r_c)
    \right)
    +
    \cdots .
    \label{4.23}
\end{equation}
Thus the monopole contribution belongs to the stationary background, whereas
the second term is the genuine quadrupolar ringdown modulation.

\section{Regime of Validity and Operational Status}
\label{sec5}

The result in Eq. \eqref{4.4} is a controlled statement about a specific
detector/cavity observable. In this section we collect the assumptions under
which the calculation is valid and clarify which parts of the modulation are
operationally meaningful.

\subsection{Scale hierarchy}
\label{ssec5.1}

The detector crosses a narrow cavity centered at \(r=r_c\) during a proper
time interval \(\Delta\tau_c\). The corresponding advanced-time width is
\begin{equation}
    \Delta v_c
    =
    \left(\frac{dv}{d\tau}\right)_{\tau_c}
    \Delta\tau_c .
    \label{5.1}
\end{equation}
The near-horizon and adiabatic approximations require
\begin{equation}
    \kappa\Delta\tau_c\ll1,
    \qquad
    |\omega|\Delta v_c\ll1,
    \qquad
    |\omega|
    =
    \sqrt{\omega_R^2+\omega_I^2}.
    \label{5.2}
\end{equation}
The first condition ensures that the Schwarzschild logarithmic redshift map
dominates over slow background variation during one transit. The second
condition ensures that the QNM factor \(e^{-i\omega v}\) is approximately
constant across the same transit, so that
\[
    \delta u(\tau)-\delta u(\tau')
    =
    (\tau-\tau')\dot{\delta u}(\tau_c)
    +
    \mathcal{O}(|\omega|^2\Delta v_c^2)
\]
can be used in the response integrals.

The single-mode and geometric-optics assumptions are
\begin{equation}
    \nu\gg\kappa,
    \qquad
    \nu\gg|\omega|,
    \qquad
    \Delta\nu\ll\nu,
    \qquad
    \nu\lesssim\omega_A .
    \label{5.3}
\end{equation}
Here \(\nu\) is the asymptotic frequency selected by the cavity-filtered
outgoing wavepacket, \(\Delta\nu\) is its bandwidth, and \(\omega_A\) is the
detector gap. Under Eq. \eqref{5.3}, corrections due to slow mode-shape
variation, state deformation, and wavepacket bandwidth are subleading and
enter the remainders in Eq. \eqref{4.5}.

Finally, the perturbative expansion requires
\begin{equation}
    0<\varepsilon\ll1,
    \qquad
    \varepsilon\,
    |\tilde{\alpha}\mathcal{C}_{20}(r_c)|
    \left(\frac{dv}{d\tau}\right)_{\tau_c}
    e^{-\omega_Iv_c}
    \ll1 .
    \label{5.4}
\end{equation}
This condition keeps the QNM-induced correction to the detailed-balance
exponent smaller than the static baseline.

For the fundamental Schwarzschild quadrupolar mode used in the figures,
\[
    \omega M
    =
    0.37367168-0.08896232\,i,
    \qquad
    |\omega|M\simeq0.384 .
\]
Thus the readout bandwidth associated with one sampling event,
\[
    B_v\equiv\frac{1}{\Delta v_c},
\]
must satisfy \(B_v\gg|\omega|\) for a clean adiabatic measurement. A
conservative choice such as
\[
    |\omega|\Delta v_c\lesssim0.1
\]
corresponds to \(B_vM\gtrsim4\). This is a sampling requirement on the
detector/cavity readout, not a requirement that the selected carrier mode be
broad: the carrier may still satisfy \(\Delta\nu\ll\nu\) provided
\begin{equation}
    |\omega|\lesssim B_v\ll\nu .
    \label{5.5}
\end{equation}

\subsection{Gauge calibration and measurable content}
\label{ssec5.2}

The eikonal perturbation is obtained from
\begin{equation}
    k^a\nabla_a\delta u
    =
    \frac12 h_{ab}k^ak^b,
    \label{5.6}
\end{equation}
where \(k^a\) is the outgoing null generator normalized by
\(k^a\nabla_a r=1\). Under an admissible EF-regular gauge transformation,
\[
    h_{ab}\mapsto h_{ab}+\nabla_a\xi_b+\nabla_b\xi_a ,
\]
the null contraction changes by an affine total derivative,
\begin{equation}
    h_{kk}
    \equiv h_{ab}k^ak^b
    \longrightarrow
    h_{kk}
    +
    2\,k^a\nabla_a(\xi_bk^b).
    \label{5.7}
\end{equation}
Consequently the primitive defining \(\delta u\), and hence
\(\mathcal{C}_{20}(r_c)\), contains an endpoint phase convention on the
cavity worldtube.

This endpoint freedom is not an additional physical degree of freedom. It is
the freedom to choose the phase origin of the cavity mode relative to the
asymptotic retarded-time phase used to define the frequency label \(\nu\).
We fix it by the calibration convention
\begin{equation}
    \Xi(r_c)=0,
    \qquad
    \Xi(r)\equiv2\xi_bk^b .
    \label{5.8}
\end{equation}
After this convention is fixed, the coefficient
\(\mathcal{C}_{20}(r_c)\) is a well-defined calibrated response coefficient
for a cavity at radius \(r_c\).

The operationally meaningful features of the modulation are then:
\begin{enumerate}
    \item the carrier frequency \(\omega_R\);
    \item the damping rate \(\omega_I\);
    \item the envelope proportional to
    \(|\mathcal{C}_{20}(r_c)|e^{-\omega_Iv_c}\);
    \item the calibrated phase offset
    \(\delta_{20}(r_c)=\arg\mathcal{C}_{20}(r_c)\);
    \item the predicted radial amplitude--phase dependence when several
    calibrated sampling radii are compared.
\end{enumerate}
A different endpoint convention would shift the absolute phase origin of the
worldtube mode, but it would not alter \(\omega_R\), \(\omega_I\), the decay
envelope, or the calibrated multi-radius consistency test. This is the sense
in which the predicted modulation is operationally meaningful within the
specified detector/cavity setup.

\subsection{Regularity, Hadamard behavior, and neglected corrections}
\label{ssec5.3}

The EF reconstruction used above is regular at the future horizon. The
ingoing QNM solution has finite metric components in ingoing
Eddington-Finkelstein coordinates, and the source \(h_{kk}\) entering
Eq. \eqref{5.6} is finite along the outgoing congruence. Therefore the
response coefficient
\[
    \mathcal{C}_{20}(r_c)
    =
    \frac12\sqrt{\frac{5}{4\pi}}
    \int_{2M}^{r_c}\mathcal{S}_{20}(r)\,dr
\]
is finite for every \(r_c>2M\). The closed boundary formula in
Sec. \ref{ssec4.2} makes this regularity explicit.

The field-theoretic short-distance behavior is also unchanged at first
order. In the Schwarzschild baseline, the outgoing-sector Wightman function
pulled back to the near-horizon detector trajectory has the standard
Hadamard singularity,
\begin{equation}
    G_0^+(s)
    =
    -\frac{\kappa^2}{16\pi}
    \frac{1}{
    \sinh^2[
    \frac{\kappa}{2}(s-i\epsilon)
    ]},
    \qquad
    s=\tau-\tau' .
    \label{5.9}
\end{equation}
The ringdown contribution enters through the smooth reparameterization
\[
    u_0(\tau)
    \longrightarrow
    u_0(\tau)+\varepsilon\delta u(\tau).
\]
Since
\[
    \delta u(\tau)-\delta u(\tau')
    =
    \mathcal{O}(\tau-\tau')
\]
near coincidence, the first-order correction does not introduce a new
ultraviolet singularity. With a smooth compact or rapidly decaying switching
function \(\chi(\tau)\), the response integrals are therefore finite.

A genuinely time-dependent geometry may also perturb the field state or the
mode profile. These effects are not part of the leading redshift-map
correction retained in Eq. \eqref{4.4}. In the present single-mode,
high-frequency regime, their size is controlled by the WKB parameters
\begin{equation}
    \mathcal{O}\!\left(
    \varepsilon\frac{\kappa}{\nu}
    \right)
    +
    \mathcal{O}\!\left(
    \varepsilon\frac{|\omega|}{\nu}
    \right)
    +
    \mathcal{O}\!\left(
    \varepsilon\frac{\Delta\nu}{\nu}
    \right).
    \label{5.10}
\end{equation}
These terms are included in the remainders of Eq. \eqref{4.5}. Thus, within
the hierarchy \eqref{5.2}--\eqref{5.4}, the dominant first-order effect is
the eikonal redshift-map deformation.

\subsection{Operational distinguishability}
\label{ssec5.4}

The time dependence of the cavity photon statistics alone is not sufficient
to identify a geometric ringdown origin. A drifting cavity resonance or a
detector detuning could also make the parameter \(2\xi(v_c)\) time
dependent. The distinguishing feature of the present effect is the QNM
template:
\begin{equation}
    \delta[2\xi]_{\rm RD}(v_c)
    \propto
    |\mathcal{C}_{20}(r_c)|
    e^{-\omega_Iv_c}
    \mathcal{S}
    \!\left(
    \omega_Rv_c-\arg\mathcal{C}_{20}(r_c)
    \right).
    \label{5.11}
\end{equation}
Thus a ringdown interpretation requires observing the expected frequency,
damping rate, and calibrated phase/amplitude dependence. In a single
uncalibrated channel, an instrumental drift artificially tuned to the same
damped sinusoid cannot be ruled out algebraically. Independent monitoring of
the cavity frequency, or comparison between multiple calibrated radii, is
therefore required to isolate the geometric contribution.

This is also the appropriate level at which to interpret possible analogue
or experimental implications. Equation \eqref{4.4} supplies a precise
template for what such a measurement would have to resolve, but the present
paper does not establish detectability in any specific platform. A realistic
implementation would need independent control of cavity detuning,
sufficient readout bandwidth \(B_v\), and a calibrated comparison with the
predicted QNM amplitude--phase structure.

\subsection{Generic content and relation to dynamical horizon thermality}
\label{ssec5.5}

It is useful to separate the content of Eq. \eqref{4.4} into generic
ringdown information and protocol-dependent detector information. The generic
part is the QNM template. In any linear, adiabatic setting in which a
near-horizon observable is sensitive to the outgoing eikonal, a ringdown
perturbation is expected to induce a correction with the characteristic
structure
\[
    e^{-\omega_I v_c}
    \left[
    A(r_c)\cos(\omega_R v_c)
    +
    B(r_c)\sin(\omega_R v_c)
    \right],
\]
or equivalently a damped sinusoid with frequency \(\omega_R\), damping rate
\(\omega_I\), and a radius-dependent phase. This follows from the linear QNM
time dependence and does not rely on the detailed microscopic model of the
two-level detector.

The protocol-dependent part is the conversion of this geometric phase
perturbation into the particular detailed-balance ratio studied here. The
single-mode projection fixes the frequency label \(\nu\); the detector gap
and switching enter through \(\tilde{\alpha}\); the cavity and wavepacket
define the bandwidth and the effective interaction time; and the endpoint
calibration fixes the absolute phase of \(\mathcal{C}_{20}(r_c)\). These
ingredients determine the magnitude, sign, and operational visibility of the
effect. They are not universal properties of the black hole alone.

This distinction also places the present work relative to broader notions of
dynamical horizon thermality. In a stationary black-hole spacetime, KMS
relations, surface gravity, and Hawking flux are mutually related in a
well-defined way. During ringdown, exact stationarity is absent, and there is
no unique global thermal state or unique time-independent temperature
associated with the perturbed geometry. Our result does not attempt to define
such a nonequilibrium temperature. Instead, it asks a narrower question:
given a stationary detector/cavity detailed-balance diagnostic, how is that
diagnostic perturbed by a controlled QNM deformation of the redshift map?
The answer is the calibrated, finite-time modulation in Eq. \eqref{4.4}.

\section{Conclusion}
\label{sec6}
We have derived a first-order ringdown correction to an operational
detector/cavity detailed-balance observable in a perturbed Schwarzschild
background. The setup is deliberately specific: a freely falling two-level
detector samples a cavity-filtered outgoing mode of fixed asymptotic
frequency \(\nu\), and the background is perturbed by an even-parity,
axisymmetric quadrupolar QNM. Within this setting, the static Schwarzschild
detailed-balance exponent remains the baseline, while the ringdown geometry
induces a controlled, decaying-oscillatory correction.

The result can be written as
\begin{equation}
    \ln
    \frac{\Gamma_{\rm abs}}{\Gamma_{\rm exc}}
    =
    \frac{2\pi\nu}{\kappa}
    -
    2\varepsilon\,
    \tilde{\alpha}\,
    |\mathcal{C}_{20}(r_c)|
    \left(\frac{dv}{d\tau}\right)_{\tau_c}
    e^{-\omega_I v_c}
    \mathcal{S}
    \!\left(
    \omega_R v_c-\delta_{20}(r_c)
    \right)
    +
    \mathcal{R},
    \label{6.1}
\end{equation}
where
\[
    \mathcal{S}(\theta)
    =
    -\omega_I\cos\theta-\omega_R\sin\theta,
    \qquad
    \delta_{20}(r_c)
    =
    \arg\mathcal{C}_{20}(r_c).
\]
Here \(\tilde{\alpha}(\nu,\omega_A,\chi)\) is a smooth detector- and
switching-dependent coefficient, while the geometric response is encoded in
\(\mathcal{C}_{20}(r_c)\). In the EF-regular reconstruction used throughout
the paper, this coefficient admits the boundary representation
\begin{equation}
    \mathcal{C}_{20}(r_c)
    =
    \frac12\sqrt{\frac{5}{4\pi}}
    \left[
    a_2(r_c)\Psi_{20}(r_c)
    +
    b_2(r_c)\partial_r\Psi_{20}(r_c)
    \right].
    \label{6.2}
\end{equation}
Thus the QNM pair \((\omega_R,\omega_I)\) determines the carrier frequency
and damping rate of the modulation, while
\(\mathcal{C}_{20}(r_c)\) fixes its calibrated radial amplitude and phase.

The correction in Eq. \eqref{6.1} is not a modification of the global
Hawking flux or of the Hawking temperature. It is a correction to a
particular windowed detector response in a single-mode, phase-calibrated,
near-horizon measurement protocol. The distinction is essential. The
frequency \(\nu\) is the frequency of the selected outgoing mode, the
coefficient \(\tilde{\alpha}\) depends on the detector gap and switching
profile, and the endpoint convention on the cavity worldtube must be fixed
before \(\mathcal{C}_{20}(r_c)\) is assigned an absolute phase. Once this
calibration is imposed, the measurable content is the QNM-template
structure: oscillation at \(\omega_R\), decay at \(\omega_I\), and the
predicted amplitude--phase dependence with sampling radius.

The result is controlled by the hierarchy
\[
    \kappa\Delta\tau_c\ll1,
    \qquad
    |\omega|\Delta v_c\ll1,
    \qquad
    \nu\gg\kappa,
    \qquad
    \nu\gg|\omega|,
    \qquad
    \Delta\nu\ll\nu,
\]
together with the perturbative bound
\[
    \varepsilon\,
    |\tilde{\alpha}\mathcal{C}_{20}(r_c)|
    \left(\frac{dv}{d\tau}\right)_{\tau_c}
    e^{-\omega_Iv_c}
    \ll1 .
\]
Under these assumptions, dynamical state or mode-shape corrections are
subleading and are included in the remainder \(\mathcal{R}\). The
Hadamard short-distance structure is unchanged at first order, because the
ringdown enters through a smooth eikonal reparameterization
\(u_0\mapsto u_0+\varepsilon\delta u\).

The static-limit checks support the interpretation of the correction as a
genuine time-dependent ringdown effect. The modulation vanishes when
\(\varepsilon\to0\), when \(v_c\to\infty\), and in the stationary quadrupolar
limit where \(\delta u\) becomes a constant phase shift. A static monopole
mass perturbation would instead change the stationary background through
\(\kappa\to\kappa_{\rm eff}\); this is a shift of the baseline rather than
the decaying-oscillatory quadrupolar signal derived here.

For a weakly leaky cavity repeatedly sampled by detector passages, the
single-mode photon statistics remain geometric but acquire a time-dependent
parameter,
\begin{equation}
    p_n(v_c)
    =
    \left[
    1-e^{-2\xi(v_c)}
    \right]
    e^{-2\xi(v_c)n},
    \qquad
    2\xi(v_c)
    =
    \ln
    \frac{\Gamma_{\rm abs}}{\Gamma_{\rm exc}} .
    \label{6.3}
\end{equation}
The induced time dependence is therefore not arbitrary: in the ringdown
scenario it must follow the damped QNM template in Eq. \eqref{6.1}. In
practice, isolating such a signal would require independent control of
cavity detuning or a comparison between several calibrated sampling radii.
A single uncalibrated channel cannot, by itself, exclude an instrumental
drift engineered to mimic the same damped sinusoid.

The extent to which these conclusions persist beyond the present assumptions
can now be stated more sharply. The damped QNM time dependence should persist
for other perturbative ringdown channels, although the response coefficient
would be replaced by the appropriate Regge-Wheeler, Zerilli, or
tensor-harmonic reconstruction coefficient, and different multipoles would
produce different angular and radial response patterns. In a slowly rotating
background, one would expect the same general separation between QNM
kinematics and detector transduction, but with Kerr QNM frequencies and
spin-\(m\)-dependent response coefficients replacing the Schwarzschild
quantities used here.

By contrast, the geometric single-mode photon statistics and the simple
closed form of Eq. \eqref{6.1} should not be assumed to survive unchanged
outside the present regime. Broader spectral filters would introduce
frequency mixing and modify \(\tilde{\alpha}\). Nonradial trajectories would
change the worldline pullback of the eikonal and the angular projection of
the perturbation. Away from the near-horizon regime, the logarithmic
Schwarzschild redshift map would no longer be the sole organizing structure
of the detector response. Beyond linear perturbation theory, mode coupling,
overtones, memory-type contributions, and nonlinear ringdown corrections
could add further time dependences that are not captured by a single
damped-sinusoid template. These extensions are natural, but they require
separate calculations and are not claimed here.

The main conclusion is therefore modest but sharp: within a controlled
near-horizon detector/cavity setup, black-hole ringdown leaves a calculable
first-order imprint on the detailed-balance exponent. The equilibrium
Schwarzschild result remains the organizing baseline, while the
time-dependent perturbation contributes a calibrated QNM correction whose
frequency, damping, and radial phase structure are fixed by the classical
ringdown geometry.

\acknowledgments
R. P. would like to acknowledge networking support of the COST Action CA21106 - COSMIC WISPers in the Dark Universe: Theory, astrophysics and experiments (CosmicWISPers), the COST Action CA22113 - Fundamental challenges in theoretical physics (THEORY-CHALLENGES), the COST Action CA21136 - Addressing observational tensions in cosmology with systematics and fundamental physics (CosmoVerse), the COST Action CA23130 - Bridging high and low energies in search of quantum gravity (BridgeQG), and the COST Action CA23115 - Relativistic Quantum Information (RQI) funded by COST (European Cooperation in Science and Technology). R. P. would also like to acknowledge the funding support of SCOAP3.

\section*{Conflict of interest}
The authors declare no conflict of interest.

\section*{Data availability statement}
Data sharing not applicable to this article as no datasets were generated or analysed during the current study.

\section*{Keywords}
Ringdown acceleration radiation; Unruh-DeWitt detector; Quasinormal modes; Linear response;
Schwarzschild black hole; Near-horizon (Rindler) limit

\appendix
\section{Switching single-mode coefficient \texorpdfstring{\(\tilde{\alpha}\)}{}, and its sign} \label{apdxA}
We collect in this appendix the definition and basic properties of the window-dependent prefactor \(\tilde{\alpha}\) that controls the first-order response of the detector in the single-mode approximation. Throughout, \(\chi(\tau)\) denotes a smooth switching function localized near the transit time \(\tau_c\) with duration \(\Delta\tau_c\). We write its autocorrelation as
\begin{equation}
    W(s)\equiv \int dT\, \chi\!\left(T+\frac{s}{2}\right)\chi\!\left(T-\frac{s}{2}\right),
    \label{A.1}
\end{equation}
which is even, nonnegative, and rapidly decaying. The perturbation enters the detector response through the windowed, single-mode coefficient \(\tilde{\alpha}\) appearing in \eqref{3.32}-\eqref{3.33}. For fixed mode frequency \(\nu\) and detector gap \(\omega_{A}\), this coefficient depends only on the switching profile \(\chi\) (equivalently on its autocorrelation \(W\)) and is independent of the ringdown perturbation. In \eqref{3.32} it multiplies \(\dot{\delta u}_{c}\), so we parameterize the first-order response by
\begin{equation}
    \frac{P_{\mathrm{exc}}-P_{\mathrm{exc}}^{(0)}}{P_{\mathrm{exc}}^{(0)}}=\varepsilon \,\dot{\delta u}_{c}\,\tilde{\alpha},\quad
    \frac{P_{\mathrm{abs}}-P_{\mathrm{abs}}^{(0)}}{P_{\mathrm{abs}}^{(0)}}=-\varepsilon \,\dot{\delta u}_{c}\,\tilde{\alpha},
    \label{A.2}
\end{equation}
The functional becomes explicit once we introduce the mean/difference parametrization \(T=\frac{1}{2}(\tau+\tau')\), \(s=\tau-\tau'\), the baseline EF-regular retarded time \(u_0(\tau)\) along the worldline, and the corresponding phase increment
\begin{equation}
    \Phi_0(s)\equiv \omega_A s - \nu\!\left[u_0\!\left(T+\frac{s}{2}\right) - u_0\!\left(T-\frac{s}{2}\right)\right].
    \label{A.3}
\end{equation}
In the adiabatic/Rindler window relevant here, \(\Phi_0\) is \(T\)-slow, so we may evaluate it at the transit mean time \(T=\tau_c\). With this simplification, the single-mode coefficient takes the compact form
\begin{equation}
    \tilde{\alpha}(\nu,\omega_A,\chi)
    = \nu\,\frac{\displaystyle \int_{-\infty}^{+\infty} ds\; s\, W(s)\, \sin\!\bigl[\Phi_0(s)\bigr]} {\displaystyle \int_{-\infty}^{+\infty} ds\; W(s)\,\bigl(1-\cos\![\Phi_0(s)]\bigr)}
    \label{A.4}
\end{equation}
with \(\Phi_0(s)\) understood at \(T=\tau_c\). The denominator equals \(P_{\mathrm{exc}}^{(0)}/\mathcal{N}\) for an appropriate normalization \(\mathcal{N}\) and is strictly positive unless the response is trivial; the expression is manifestly finite because \(W\) decays and \(\Phi_0(s)=\mathcal{O}(s)\) near \(s=0\). The overall sign convention for the first-order correction is fixed elsewhere in the main text (see Theorem 1), so \(\tilde{\alpha}\) can be taken as the window-controlled prefactor.

We can read off several consequences directly from \eqref{A.4}. Since \(W\) is even and \(\Phi_0\) is odd in \(s\), the denominator is nonnegative and vanishes only in the absence of a transition, whereas the numerator has the same sign as the local linear coefficient of \(\Phi_0\). Writing
\begin{equation}
    \phi_1 \equiv \omega_A - \nu\,\dot u_0(\tau_c),
\end{equation}
we obtain, in the narrow-window regime \(\Delta\tau_c \to 0\), the uniform expansion \(\sin\Phi_0(s)=\phi_1 s+\mathcal{O}(s^3)\) and \(1-\cos\Phi_0(s)=\frac{1}{2}\phi_1^2 s^2 + \mathcal{O}(s^4)\). Substituting into \eqref{A.4} and using \(\int s\,W(s)\,ds=0\) and \(\int s^2 W(s)\,ds>0\), we arrive at
\begin{equation}
    \tilde{\alpha}
    = \nu\,\frac{\phi_1\int s^2 W(s)\,ds}{\frac{1}{2}\phi_1^2\int s^2 W(s)\,ds}
    \;=\; \frac{2\nu}{\phi_1} \;+\; \mathcal{O}(\Delta\tau_c^{2}),
    \quad
    \phi_1=\omega_A-\nu\,\dot u_0(\tau_c),
    \label{A.5}
\end{equation}
showing that \(\tilde{\alpha}=\mathcal{O}(1)\) as \(\Delta\tau_c\to 0\), with sign \(\mathrm{sgn}(\tilde{\alpha})=\mathrm{sgn}(\phi_1)\) and magnitude set solely by \(\phi_1\) at leading order, independent of the detailed shape of \(W\).

The sign change of \(\tilde{\alpha}\) has a simple physical interpretation.
It occurs when the detector gap is crossed by the local proper-time frequency
of the selected outgoing mode. Indeed, for the radial \(E=1\) trajectory,
\begin{equation}
    \dot u_0(r_c)
    =
    \frac{dt}{d\tau}
    -
    \frac{dr_*}{dr}\frac{dr}{d\tau}
    =
    \frac{1+\sqrt{2M/r_c}}{1-2M/r_c},
    \label{A.5a}
\end{equation}
so the leading-order sign flip occurs at
\begin{equation}
    \omega_A=\nu\,\dot u_0(r_c).
    \label{A.5b}
\end{equation}
Equivalently, for fixed detector gap and selected cavity frequency, this is a
condition on the sampling radius. In the near-horizon limit,
\[
    \dot u_0(r_c)\simeq \frac{4M}{r_c-2M},
\]
and therefore
\begin{equation}
    r_c-2M
    \simeq
    4M\,\frac{\nu}{\omega_A},
    \qquad
    \left(\frac{\omega_A}{\nu}\gg1\right).
    \label{A.5c}
\end{equation}
Thus the sign flip is a detector-resonance effect rather than a separate
geometric feature of the ringdown perturbation. For generic off-resonant
choices of \((\omega_A,\nu,r_c)\), the sign of \(\tilde{\alpha}\) remains
fixed. The apparent singularity in the narrow-window expression
\(\tilde{\alpha}\simeq 2\nu/\phi_1\) is an artifact of taking
\(\Delta\tau_c\to0\) before resolving the resonance; for any finite switching
time the exact expression \eqref{A.4} remains finite, with the response
regularized by the finite bandwidth of the switching function.

It is also useful to note what changes if one replaces the single selected
frequency by a smooth spectral distribution. Let \(w(\nu)\) denote a positive
frequency weight, for example a thermal occupation or a cavity-filtered
thermal spectrum, normalized over the frequency band sampled by the detector.
At leading order in the same adiabatic/geometric-optics regime used in the
main text, the QNM-dependent time factor is independent of \(\nu\). Hence a
spectral average modifies only the effective detector prefactor,
\begin{equation}
    \tilde{\alpha}_{\rm eff}
    =
    \frac{\displaystyle\int_0^\infty d\nu\,
    w(\nu)\,\tilde{\alpha}(\nu,\omega_A,\chi)}
    {\displaystyle\int_0^\infty d\nu\,w(\nu)} ,
    \label{A.5d}
\end{equation}
so that the averaged modulation retains the form
\begin{equation}
    \delta\!\left[
    \ln\frac{\Gamma_{\rm abs}}{\Gamma_{\rm exc}}
    \right]_{\rm av}
    =
    -2\varepsilon\,\tilde{\alpha}_{\rm eff}
    |\mathcal{C}_{20}(r_c)|
    \left(\frac{dv}{d\tau}\right)_{\tau_c}
    e^{-\omega_I v_c}
    \mathcal{S}\!\left(\omega_R v_c-\delta_{20}(r_c)\right),
    \label{A.5e}
\end{equation}
up to the same adiabatic and narrow-band corrections appearing in
Theorem 1. Therefore, a thermal or otherwise broadband average can suppress
the amplitude if the weighting samples both sides of the detuning
\(\omega_A-\nu\dot u_0(r_c)=0\), but it does not dephase the ringdown
oscillation itself at leading order. Complete cancellation would require a
special weighting that balances the positive- and negative-detuning
contributions. The single-mode formula used in the main text is therefore
the spectrally resolved limit of this more general averaged expression.

In the opposite, wide-window (adiabatic) regime, both integrals in \eqref{A.4} are suppressed by stationary-phase/steepest-descent arguments, but with the same powers of \(\Delta\tau_c\); the ratio therefore remains finite and \(\mathcal{O}(1)\) across the full adiabatic window. Using the elementary bounds \(|\sin \Phi_0|\le |\Phi_0|\) and \(1-\cos \Phi_0 \ge \frac{2}{\pi^2}\min\{\Phi_0^2,\pi^2\}\), we also obtain the uniform estimate
\begin{equation}
    0 \;\le\; |\tilde{\alpha}|
    \;\le\;
    \nu\,
    \frac{\displaystyle \int |s|\,W(s)\,|\Phi_0(s)|\,ds}
         {\displaystyle \int \frac{2}{\pi^2}\min\{\Phi_0(s)^2,\pi^2\}\,W(s)\,ds}
    \;=\; \mathcal{O}(1),
    \label{A.6}
\end{equation}
valid in the regime specified in the main text, with constants depending only on the fixed shape of \(W\).

For practical use, it is convenient to have closed forms for standard windows. If \(\chi(\tau)=\chi_0\exp\!\left[-(\tau-\tau_c)^2/(2\Delta\tau_c^2)\right]\) is Gaussian, then
\begin{equation}
    W(s)=W_0\,\exp\!\left[-\frac{s^2}{4\Delta\tau_c^{2}}\right],
    \quad
    W_0=\sqrt{\pi}\,\Delta\tau_c\,\chi_0^2,
    \label{A.7}
\end{equation}
and \eqref{A.4} becomes the ratio of one-dimensional oscillatory integrals
\begin{equation}
    \tilde{\alpha}_{\mathrm{G}}(\nu,\omega_A,\Delta\tau_c)
    =
    \nu\,
    \frac{\displaystyle \int_{-\infty}^{+\infty} ds\; s\,e^{-s^{2}/(4\Delta\tau_c^{2})}\,\sin\!\bigl[\Phi_0(s)\bigr]}
         {\displaystyle \int_{-\infty}^{+\infty} ds\; e^{-s^{2}/(4\Delta\tau_c^{2})}\,\bigl(1-\cos\!\left[\Phi_0(s)\right]\bigr)}.
    \label{A.8}
\end{equation}
The small-\(\Delta\tau_c\) series follows immediately from \eqref{A.5}; higher-order corrections can be organized in terms of Hermite-Gaussian moments of \(\sin\Phi_0\) and \(\cos\Phi_0\).

If, instead, \(\chi(\tau)=\chi_0\, b\!\left(\frac{\tau-\tau_c}{\Delta\tau_c}\right)\) is a compactly supported \(C^\infty\) bump with \(b\in C_0^\infty([-1,1])\), \(b\) even, and \(\int b^2=1\), then
\begin{equation}
    W(s)=\Delta\tau_c\,\chi_0^2\,w\!\left(\frac{s}{\Delta\tau_c}\right),
    \quad
    w(\sigma)=\int_{-1}^{+1} d\eta\; b\!\left(\eta+\frac{\sigma}{2}\right)b\!\left(\eta-\frac{\sigma}{2}\right),
\end{equation}
with \(w\) even and supported in \(|\sigma|\le2\). Writing \(\Phi_0(\Delta\tau_c,\sigma)\equiv \Phi_0(s=\Delta\tau_c \sigma)\), we obtain the compact representation
\begin{equation}
    \tilde{\alpha}_{\mathrm{B}}(\nu,\omega_A,\Delta\tau_c)
    =
    \nu\,
    \frac{\displaystyle \int_{-2}^{+2} d\sigma\; \sigma\, w(\sigma)\,\sin\!\bigl[\Phi_0(\Delta\tau_c,\sigma)\bigr]}
         {\displaystyle \int_{-2}^{+2} d\sigma\; w(\sigma)\,\bigl(1-\cos\![\Phi_0(\Delta\tau_c,\sigma)]\bigr)},
    \label{A.10}
\end{equation}
from which the narrow-window asymptotics again reproduces \eqref{A.5}. In both cases, the \(T\)-slow dependence enters solely through \(\Phi_0\) evaluated at \(T=\tau_c\) via the near-horizon EF map used in the main text.

\section{EF-regular reconstruction and boundary formula for
\texorpdfstring{\(\mathcal{C}_{20}(r_c)\)}{C20}}
\label{apdxB}

This appendix records the EF-regular reconstruction needed to obtain the
boundary response coefficient \(\mathcal{C}_{20}(r_c)\). The purpose is not
to reproduce the full even-parity perturbation formalism, but to isolate the
single metric combination that enters the eikonal transport equation,
\[
    k^a\nabla_a\delta u
    =
    \frac12 h_{ab}k^ak^b .
\]
Throughout this appendix we work in ingoing Eddington-Finkelstein
coordinates \((v,r,\theta,\phi)\), with
\[
    f(r)=1-\frac{2M}{r},
    \qquad
    u_0=v-2r_*(r),
\]
and use the outgoing null generator
\begin{equation}
    k^a
    =
    -g_{(0)}^{ab}\nabla_b u_0
    =
    \left(\partial_r\right)^a
    +
    \frac{2}{f(r)}
    \left(\partial_v\right)^a .
    \label{B.1}
\end{equation}
This normalization gives
\[
    k^a\nabla_a u_0=0,
    \qquad
    k^a\nabla_a r=1,
\]
so the areal radius \(r\) is an affine parameter along the background
outgoing rays. Hence
\[
    k^a\nabla_a
    =
    \frac{d}{dr}
\]
when acting along the outgoing congruence.

We consider the even-parity \((\ell,m)=(2,0)\) perturbation written in terms
of the Zerilli-Moncrief master field,
\[
    \Psi_{20}(v,r)
    =
    \hat\Psi_{20}(r,\omega)e^{-i\omega v}.
\]
On the symmetry axis, \(Y_{20}(0)=\sqrt{5/(4\pi)}\), and all angular
derivatives vanish. The only metric datum entering the eikonal correction is
the null-null contraction
\[
    h_{kk}
    \equiv
    h_{ab}k^ak^b .
\]
For an EF-regular reconstruction, this contraction can be placed in the
radial divergence form
\begin{equation}
    h_{kk}(v,r,\theta=0)
    =
    \Re\!\left\{
    \frac{d}{dr}
    \left[
    a_2(r)\hat\Psi_{20}(r)
    +
    b_2(r)\partial_r\hat\Psi_{20}(r)
    \right]
    e^{-i\omega v}
    \right\}.
    \label{B.2}
\end{equation}
The functions \(a_2(r)\) and \(b_2(r)\) are regular at the future horizon.
For \(\ell=2\), with
\[
    \Lambda(r)=2r+3M,
\]
one convenient EF-regular choice is
\begin{equation}
    a_2(r)
    =
    \frac{
    (r-2M)(2r^2+6Mr+6M^2)
    }{
    r^2\Lambda(r)^2
    },
    \qquad
    b_2(r)
    =
    \frac{
    (r-2M)^2
    }{
    r\Lambda(r)^2
    } .
    \label{B.3}
\end{equation}
These profiles vanish at the horizon as
\begin{equation}
    a_2(r)=\mathcal{O}(r-2M),
    \qquad
    b_2(r)=\mathcal{O}\!\left((r-2M)^2\right),
    \qquad
    r\to2M .
    \label{B.4}
\end{equation}
Thus the primitive in Eq. \eqref{B.2} is finite for an ingoing
EF-regular QNM solution.

The eikonal perturbation at the sampling radius follows by integrating
\[
    \frac{d}{dr}\delta u
    =
    \frac12 h_{kk}
\]
from the future horizon to \(r=r_c\). Extracting the harmonic factor, one
obtains
\begin{equation}
    \delta u(r_c,v_c)
    =
    \Re\!\left[
    \mathcal{C}_{20}(r_c)e^{-i\omega v_c}
    \right],
    \label{B.5}
\end{equation}
with
\begin{equation}
    \mathcal{C}_{20}(r_c)
    =
    \frac12
    \sqrt{\frac{5}{4\pi}}
    \left[
    a_2(r_c)\hat\Psi_{20}(r_c)
    +
    b_2(r_c)\partial_r\hat\Psi_{20}(r_c)
    \right].
    \label{B.6}
\end{equation}
This is the boundary formula used in the main text. Equivalently, if the
overall ringdown amplitude is written explicitly as
\[
    \Psi_{20}(v,r)
    =
    \mathcal{A}_{20}\psi_2(r)e^{-i\omega v},
\]
then
\begin{equation}
    \mathcal{C}_{20}(r_c)
    =
    \frac12
    \sqrt{\frac{5}{4\pi}}
    \mathcal{A}_{20}
    \left[
    a_2(r_c)\psi_2(r_c)
    +
    b_2(r_c)\psi_2'(r_c)
    \right].
    \label{B.7}
\end{equation}

The same expression may be written as the integral response
\begin{equation}
    \mathcal{C}_{20}(r_c)
    =
    \frac12
    \sqrt{\frac{5}{4\pi}}
    \int_{2M}^{r_c}\mathcal{S}_{20}(r)\,dr ,
    \label{B.8}
\end{equation}
where \(\mathcal{S}_{20}(r)\) is the EF-regular radial source defined by
\[
    h_{kk}
    =
    \Re\!\left[
    \mathcal{S}_{20}(r)e^{-i\omega v}
    \right] .
\]
Equations \eqref{B.2} and \eqref{B.6} show that this integral reduces to a
boundary term. In particular,
\begin{equation}
    \mathcal{S}_{20}(r)
    =
    \frac{d}{dr}
    \left[
    a_2(r)\hat\Psi_{20}(r)
    +
    b_2(r)\partial_r\hat\Psi_{20}(r)
    \right].
    \label{B.9}
\end{equation}

Several consistency checks follow immediately. Near the horizon,
Eq. \eqref{B.4} implies
\begin{equation}
    \mathcal{C}_{20}(r_c)
    =
    \frac12
    \sqrt{\frac{5}{4\pi}}
    a_2'(2M)\hat\Psi_{20}(2M)(r_c-2M)
    +
    \mathcal{O}\!\left((r_c-2M)^2\right),
    \label{B.10}
\end{equation}
so the response grows linearly with the cavity altitude above the horizon.
This is the behavior displayed in Fig. \ref{fig1}.

In the far zone, the QNM solution has the outgoing asymptotic behavior
\[
    \hat\Psi_{20}(r)
    \sim
    A_{\rm out}\,r^{-1}e^{i\omega r_*}
\]
up to the chosen normalization. Since \(a_2(r)\) and \(b_2(r)\) have
regular large-\(r\) expansions, Eq. \eqref{B.6} gives the expected
radiative falloff of \(\mathcal{C}_{20}(r_c)\), modulo the oscillatory QNM
phase. This is the large-radius behavior used in interpreting the radial
profile of Fig. \ref{fig1}.

The boundary expression is unique up to additions proportional to the
Zerilli equation or to residual endpoint rephasings of the cavity
worldtube. The former vanish on solutions of the master equation. The latter
are the calibration freedom discussed in Appendix \ref{apdxD}; after the
endpoint convention \(\Xi(r_c)=0\) is imposed, Eq. \eqref{B.6} defines the
calibrated response coefficient used in the detailed-balance modulation.

\section{Numerical evaluation of \texorpdfstring{\(\mathcal{C}_{20}(r_c)\)}{C20}}
\label{apdxC20num}

This appendix records the numerical implementation used for
Fig. \ref{fig1} and provides the convergence checks for the response
coefficient \(\mathcal{C}_{20}(r_c)\). The purpose is only to document the
evaluation of the EF-regular boundary formula; the analytic result in the
main text does not depend on a particular normalization of the radial QNM.

For the fundamental Schwarzschild even-parity mode
\((\ell,m,n)=(2,0,0)\), we use
\[
    \omega M = 0.37367168-0.08896232\,i
\]
with the Fourier convention
\[
    \Psi_{20}(v,r)=\hat\Psi_{20}(r,\omega)e^{-i\omega v}.
\]
The EF-regular radial function satisfies
\[
    f(r)\,\hat\Psi_{20}''(r)
    +
    \bigl(f'(r)-2i\omega\bigr)\hat\Psi_{20}'(r)
    -
    V^{\rm Z}_{2}(r)\hat\Psi_{20}(r)
    =
    0,
    \qquad
    f(r)=1-\frac{2M}{r},
\]
where \(V^{\rm Z}_{2}\) is the Zerilli potential evaluated at \(\ell=2\).
The horizon-regular normalization is chosen as
\[
    \hat\Psi_{20}(2M)=1.
\]
EF regularity then gives
\[
    \hat\Psi_{20}'(2M)=0,
\]
and the near-horizon Frobenius expansion begins as
\[
    \hat\Psi_{20}(r)
    =
    1+
    \frac{189}{784\,M^{2}}
    \frac{(r-2M)^2}{1-2i\omega M}
    +
    \mathcal{O}\!\left((r-2M)^3\right).
\]
In practice, the integration is started at
\[
    r_{\rm in}=2M(1+10^{-6})
\]
with the above regular initial data and then evolved outward using an
adaptive Runge--Kutta method. The response coefficient is evaluated from the
closed boundary expression
\[
    \mathcal{C}_{20}^{\rm bd}(r_c)
    =
    \frac12\sqrt{\frac{5}{4\pi}}
    \left[
    a_2(r_c)\hat\Psi_{20}(r_c)
    +
    b_2(r_c)\partial_r\hat\Psi_{20}(r_c)
    \right],
\]
with \(a_2(r)\) and \(b_2(r)\) given in Sec. \ref{ssec4.2}. Since
\(\mathcal{C}_{20}\propto\mathcal{A}_{20}\), Fig. \ref{fig1} displays the
curves in arbitrary normalization.

\begin{figure}
    \centering
    \includegraphics[width=0.6\textwidth]{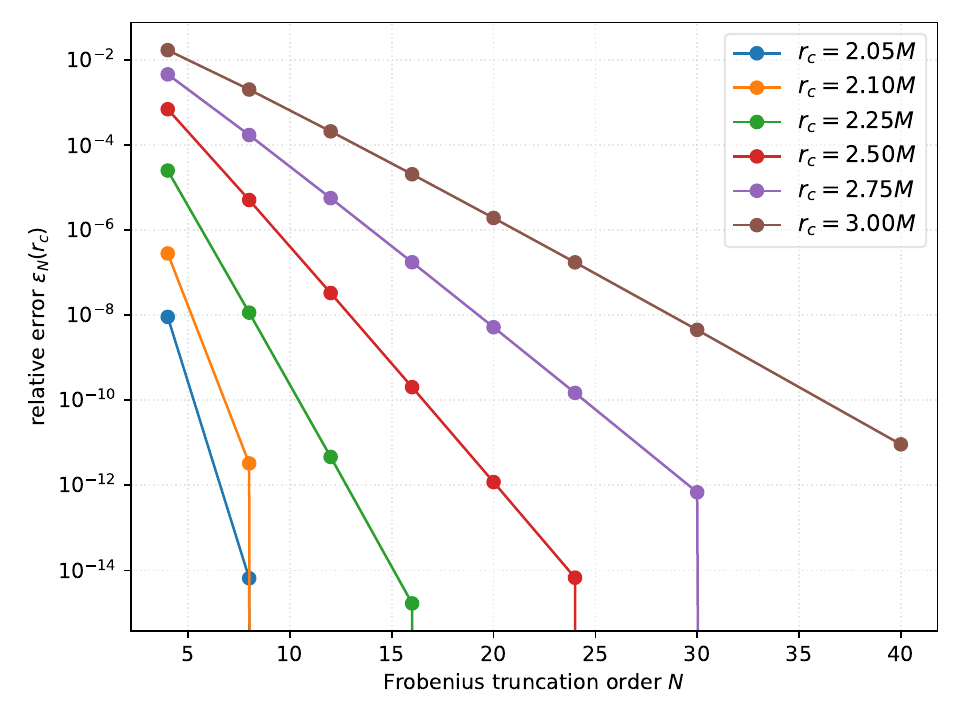}
    \caption{
    Relative truncation error \(\epsilon_N(r_c)\) in
    \(|\mathcal{C}_{20}(r_c)|\), computed from the truncated EF-regular
    Frobenius expansion and the closed boundary formula. The rapid
    convergence in the near-horizon interval confirms that the behavior in
    Fig. \ref{fig1} is not a Frobenius-truncation artifact.
    }
    \label{fig:C20convergence}
\end{figure}

As a convergence check near the horizon, we also compute
\(\mathcal{C}_{20}(r_c)\) using a truncated Frobenius solution,
\[
    \hat\Psi_{20}^{(N)}(r,\omega)
    =
    \sum_{n=0}^{N}c_n(r-2M)^n,
    \qquad
    c_0=1,\qquad c_1=0,
\]
where the coefficients \(c_n\) are obtained recursively from the radial
equation. Defining
\[
    \mathcal{C}_{20}^{(N)}(r_c)
    =
    \frac12\sqrt{\frac{5}{4\pi}}
    \left[
    a_2(r_c)\hat\Psi_{20}^{(N)}(r_c)
    +
    b_2(r_c)\partial_r\hat\Psi_{20}^{(N)}(r_c)
    \right],
\]
we monitor the relative truncation error
\[
    \epsilon_N(r_c)
    =
    \frac{
    \left|
    |\mathcal{C}_{20}^{(N)}(r_c)|
    -
    |\mathcal{C}_{20}^{(N_{\rm ref})}(r_c)|
    \right|}
    {
    |\mathcal{C}_{20}^{(N_{\rm ref})}(r_c)|
    } .
\]
As shown in Fig. \ref{fig:C20convergence}, the convergence is rapid
throughout the near-horizon interval \(2M<r_c\lesssim3M\). This is expected
from the boundary formula, because near \(r=2M\),
\[
    a_2(r)=\mathcal{O}(r-2M),
    \qquad
    b_2(r)=\mathcal{O}\!\left((r-2M)^2\right),
\]
so the leading contribution is controlled by the regular horizon value of
\(\hat\Psi_{20}\).

Finally, we checked the boundary formula against the direct transport
integral. The integral evaluation is
\[
    \mathcal{C}_{20}^{\rm int}(r_c)
    =
    \frac12\sqrt{\frac{5}{4\pi}}
    \int_{2M}^{r_c}\mathcal{S}_{20}(r)\,dr,
\]
while the boundary evaluation is
\[
    \mathcal{C}_{20}^{\rm bd}(r_c)
    =
    \frac12\sqrt{\frac{5}{4\pi}}
    \left[
    a_2(r_c)\hat\Psi_{20}(r_c)
    +
    b_2(r_c)\partial_r\hat\Psi_{20}(r_c)
    \right].
\]
We monitor
\[
    \Delta_C(r_c)
    =
    \frac{
    \left|
    \mathcal{C}_{20}^{\rm int}(r_c)
    -
    \mathcal{C}_{20}^{\rm bd}(r_c)
    \right|
    }{
    \left|
    \mathcal{C}_{20}^{\rm bd}(r_c)
    \right|
    } .
\]
Across the range of sampling radii used in Fig. \ref{fig1}, the two
evaluations agree within the numerical tolerance of the radial integration
and quadrature. This confirms the total-derivative reconstruction used in
the boundary expression.

\section{Gauge calibration and endpoint phase convention}
\label{apdxD}

This appendix clarifies the residual gauge freedom associated with the
eikonal correction
\[
    u=u_0+\varepsilon\,\delta u
\]
and with the response coefficient \(\mathcal{C}_{20}(r_c)\). The essential
point is that \(\mathcal{C}_{20}(r_c)\) is a calibrated worldtube response:
its absolute phase is defined only after the phase of the cavity mode has
been fixed relative to the asymptotic retarded-time coordinate used to label
the outgoing frequency \(\nu\).

At linear order, an infinitesimal gauge transformation generated by
\(\xi^a\) acts as
\begin{equation}
    h_{ab}
    \longrightarrow
    h'_{ab}
    =
    h_{ab}
    +
    \nabla_a\xi_b
    +
    \nabla_b\xi_a .
    \label{D.1}
\end{equation}
We restrict to admissible transformations that are regular in ingoing
Eddington-Finkelstein coordinates and preserve the normalization of the
outgoing generator
\[
    k^a\nabla_a r=1 .
\]
Since \(k^a\) is affinely parametrized by \(r\), we have
\[
    k^a\nabla_a k^b=0,
    \qquad
    k^a\nabla_a=\frac{d}{dr}
\]
along the background outgoing congruence. Contracting Eq. \eqref{D.1} twice
with \(k^a\) gives
\begin{equation}
    h_{kk}
    \equiv
    h_{ab}k^ak^b
    \longrightarrow
    h'_{kk}
    =
    h_{kk}
    +
    2k^a\nabla_a(\xi_bk^b)
    =
    h_{kk}
    +
    \frac{d}{dr}\Xi(r),
    \label{D.2}
\end{equation}
where
\begin{equation}
    \Xi(r)
    \equiv
    2\,\xi_bk^b .
    \label{D.3}
\end{equation}
Thus an admissible EF-regular gauge transformation changes the source in the
eikonal transport equation only by an affine total derivative.

The eikonal correction satisfies
\begin{equation}
    \frac{d}{dr}\delta u
    =
    \frac12 h_{kk}.
    \label{D.4}
\end{equation}
Therefore Eq. \eqref{D.2} implies
\begin{equation}
    \delta u(r_c)
    \longrightarrow
    \delta u(r_c)
    +
    \frac12
    \left[
    \Xi(r_c)-\Xi(2M)
    \right].
    \label{D.5}
\end{equation}
For regular ingoing perturbations, the horizon value may be fixed by the
choice of horizon phase,
\begin{equation}
    \Xi(2M)=0 .
    \label{D.6}
\end{equation}
The remaining freedom is then the endpoint value \(\Xi(r_c)\) on the cavity
worldtube.

Equivalently, since
\begin{equation}
    \delta u(r_c,v_c)
    =
    \Re\!\left[
    \mathcal{C}_{20}(r_c)e^{-i\omega v_c}
    \right],
    \label{D.7}
\end{equation}
the response coefficient transforms as
\begin{equation}
    \mathcal{C}_{20}(r_c)
    \longrightarrow
    \mathcal{C}'_{20}(r_c)
    =
    \mathcal{C}_{20}(r_c)
    +
    \frac12
    \sqrt{\frac{5}{4\pi}}\,
    \Xi(r_c),
    \label{D.8}
\end{equation}
after the \(Y_{20}(0)=\sqrt{5/(4\pi)}\) factor has been extracted. This is
the endpoint ambiguity of the boundary primitive. It is not an additional
physical perturbation; it is the freedom to choose the phase origin of the
cavity mode on the worldtube.

The detector couples to the outgoing phase \(e^{-i\nu u}\). On a finite
worldtube \(r=r_c\), shifting
\[
    u
    \longrightarrow
    u+\varepsilon\,\Xi(r_c)
\]
amounts to the rephasing
\begin{equation}
    e^{-i\nu u}
    \longrightarrow
    e^{-i\nu u}
    e^{-i\varepsilon\nu\Xi(r_c)} .
    \label{D.9}
\end{equation}
This constant phase can be absorbed into the definition of the cavity-mode
operator. To compare the cavity mode with the outgoing frequency label
\(\nu\) defined at \(\mathscr{I}^+\), however, one must choose a phase
convention. In this paper we impose the endpoint calibration
\begin{equation}
    \Xi(r_c)=0 .
    \label{D.10}
\end{equation}
With this convention fixed, \(\mathcal{C}_{20}(r_c)\) and its phase
\[
    \delta_{20}(r_c)=\arg\mathcal{C}_{20}(r_c)
\]
are the calibrated amplitude and phase of the detector response at the
sampling radius.

This calibration does not affect the QNM-template content of the prediction.
A different constant endpoint convention would shift the absolute phase
origin of the worldtube mode, but it would not change the QNM frequency
\(\omega_R\), the damping rate \(\omega_I\), or the envelope proportional to
\[
    |\mathcal{C}_{20}(r_c)|e^{-\omega_Iv_c}
\]
once the same convention is used consistently.

For several cavities at radii \(r_i\), the calibration must be imposed
relative to a common asymptotic retarded-time standard. The natural
multi-worldtube convention is
\begin{equation}
    \Xi(r_i)=0,
    \qquad
    i=1,\ldots,N .
    \label{D.11}
\end{equation}
Then the radial dependence of
\(\mathcal{C}_{20}(r_i)\) is operationally meaningful. In particular, the
ringdown interpretation predicts that all calibrated channels share the same
\((\omega_R,\omega_I)\), while their relative amplitudes and phases are fixed
by
\[
    |\mathcal{C}_{20}(r_i)|,
    \qquad
    \arg\mathcal{C}_{20}(r_i).
\]
Without a common calibration, independent endpoint rephasings could obscure
these relative phases.

Thus the gauge statement used in the main text is the following. The local
source \(h_{kk}\) changes by a total derivative under admissible EF-regular
gauge transformations. The corresponding primitive
\(\mathcal{C}_{20}(r_c)\) has an endpoint phase ambiguity. Once the cavity
worldtube phase is calibrated by \(\Xi(r_c)=0\), the detailed-balance
modulation in Eq. \eqref{4.4} is a well-defined detector/cavity observable
within the specified operational setup.

\section{Notation and limiting conventions}
\label{apdxE}

This appendix summarizes the notation and limiting conventions used in the
main text. It is included to avoid ambiguity in the interpretation of the
response coefficient, the QNM phase convention, and the static limits.

We use ingoing Eddington-Finkelstein coordinates \((v,r,\theta,\phi)\), with
\[
    v=t+r_*,
    \qquad
    u_0=v-2r_*(r),
    \qquad
    f(r)=1-\frac{2M}{r}.
\]
The background outgoing null generator is normalized as
\[
    k^a
    =
    -g_{(0)}^{ab}\nabla_bu_0,
    \qquad
    k^a\nabla_a r=1.
\]
Thus \(r\) is an affine parameter along the outgoing congruence and
\[
    k^a\nabla_a=\frac{d}{dr}
\]
when acting along these background rays.

The QNM convention is
\[
    \Psi_{20}(v,r)
    =
    \hat\Psi_{20}(r,\omega)e^{-i\omega v},
    \qquad
    \omega=\omega_R-i\omega_I,
    \qquad
    \omega_I>0 .
\]
With this convention,
\[
    e^{-i\omega v}
    =
    e^{-i\omega_R v}e^{-\omega_I v},
\]
so the QNM contribution decays for increasing \(v\).

The eikonal perturbation at the sampling radius is written as
\[
    \delta u(r_c,v_c)
    =
    \Re\!\left[
    \mathcal{C}_{20}(r_c)e^{-i\omega v_c}
    \right],
\]
where \(\mathcal{C}_{20}(r_c)\) is the calibrated response coefficient.
After the endpoint phase convention \(\Xi(r_c)=0\) is imposed, its magnitude
and phase determine the amplitude and phase of the detector response:
\[
    \mathcal{C}_{20}(r_c)
    =
    |\mathcal{C}_{20}(r_c)|e^{i\delta_{20}(r_c)},
    \qquad
    \delta_{20}(r_c)=\arg\mathcal{C}_{20}(r_c).
\]
A different endpoint phase convention would change the absolute phase
assigned to the cavity mode, but not the QNM frequency, damping rate, or the
calibrated response once a common convention is used.

The phrase \textit{near-horizon limit} always means \(r_c\to2M^+\) at fixed
Schwarzschild mass \(M\), with the detector still sampling outside the
future horizon. In this limit, the EF-regular boundary formula gives
\[
    \mathcal{C}_{20}(r_c)
    =
    \mathcal{O}(r_c-2M),
\]
because the coefficient \(a_2(r)\) is \(\mathcal{O}(r-2M)\) and
\(b_2(r)\) is \(\mathcal{O}((r-2M)^2)\).

The phrase \textit{late-time limit} means
\[
    v_c\to+\infty
\]
at fixed perturbation amplitude and fixed sampling radius. Since
\(\omega_I>0\), the QNM factor \(e^{-\omega_Iv_c}\) vanishes and the
ringdown correction to the detailed-balance exponent disappears.

The phrase \textit{stationary quadrupolar limit} refers to the limit in which the
quadrupolar eikonal perturbation becomes time independent. In that case
\[
    \delta u(\tau)
    =
    \text{constant}
    +
    \mathcal{O}(\omega),
\]
and the first-order correction to the probability ratio vanishes because
only the antisymmetric difference
\[
    \delta u(\tau)-\delta u(\tau')
\]
enters the detector response. A constant shift of \(u\) is therefore only a
constant rephasing of the selected cavity mode.

This stationary quadrupolar limit should not be confused with a monopole
mass perturbation. A static monopole perturbation changes the background
mass and hence the surface gravity,
\[
    M\to M+\varepsilon\delta M,
    \qquad
    \kappa\to\kappa_{\rm eff}
    =
    \frac{1}{4(M+\varepsilon\delta M)}.
\]
Such a contribution changes the stationary detailed-balance baseline. It is
not the decaying-oscillatory quadrupolar ringdown signal studied in this
work.

\bibliography{ref}

\end{document}